\documentclass[prd,aps,preprintnumbers,floats,floatfix,superscriptaddress,nofootinbib,notitlepage,twocolumn]{revtex4} 

\usepackage{amsmath}  
\usepackage{amssymb}
\usepackage{amsfonts} 
\usepackage{amssymb}  
\usepackage{natbib}     
\usepackage{graphicx}
\graphicspath {  { C:\Users\Laura\Desktop\PNScooling\paper1_plots }  }
\usepackage{dcolumn}
\usepackage{bm}
\usepackage{caption}
\usepackage{subcaption}
\usepackage{subcaption} 
\usepackage{ragged2e}
\justifying
\usepackage{tabularx} 
\usepackage{array}
\usepackage[maxfloats=256]{morefloats}
\usepackage{comment}
\usepackage[colorlinks=true, allcolors=blue]{hyperref}

\begin{document}


\title{Asymmetric emissions of neutrinos in the cooling of rotating proto-neutron stars}

\author{Laura \sc{Barrio}}
\email{laura@heap.phys.waseda.ac.jp}
\affiliation{Graduate School of Advanced Science and Engineering, Waseda University, 3-4-1 Okubo, Shinjuku, Tokyo 169-8555, Japan.}

\author{Kotaro \sc{Fujisawa}}
\affiliation{
Department of Liberal Arts, Tokyo University of Technology, Ota-ku, Tokyo 144-0051, Japan
}

\author{Ryuichiro \sc{Akaho}}
\affiliation{
Faculty of Science and Engineering, Waseda University, 3-4-1 Okubo, Shinjuku, Tokyo 169-8555, Japan}

\author{Hiroki \sc{Nagakura}}
\affiliation{
National Astronomical Observatory of Japan, 2-21-1 Osawa, Mitaka, Tokyo 181-8588, Japan
}

\author{Shoichi \sc{Yamada}}
\affiliation{
Advanced Research Institute for Science and Engineering, Waseda University, 3-4-1 Okubo, Shinjuku, Tokyo 169-8555, Japan
}

\date{\today}

\begin{abstract}
We evaluate global asymmetry in the luminosities of neutrinos emitted from rapidly-rotating proto-neutron stars (PNS's). We build axisymmetric models of PNS's in mechanical equilibrium with rotation by adding prescribed angular momentum distributions by hand to non-rotational PNS models, which are extracted from a one-dimensional (spherically symmetric) PNS cooling calculation at different times: \(t=2, 6, 10, 20, 30\)s after a supernova explosion. We then conduct two-dimensional (spatially axisymmetric) neutrino transport calculations on top of them with the matter profiles (and the spacetime geometry) fixed. We find for the rapidly-rotating models with  \(T/|W|\sim  5\times 10^{-2}\) that the neutrino luminosity changes by \(\sim 3 \% \) depending on the observer position. We give detailed analyses of the neutrino-hemispheres as well as the neutrino luminosities that are defined observer-wise. We also calculate the low-frequency (\(\lesssim 1{\rm Hz}\)) gravitational waves produced by the neutrinos radiated asymmetrically. We find that those gravitational waves, if emitted from the Galactic center, can be detected by planned detectors such as B-DECIGO, DECIGO and AILA. Finally, we look for crossings in the energy-integrated angular distributions in momentum space for the electron neutrino sector, a signature of the fast flavor conversion. We find them near the PNS surface in all models.
\end{abstract}


\maketitle

\section{Introduction}\label{Intro}

The fate of a star is mainly determined by its mass: while low mass stars (\(\lesssim 8M_\odot\)) end their lives as white dwarfs, high mass stars (\(\gtrsim8M_\odot\)) explode violently as core-collapse supernovae (CCSNe) and eventually leave neutron stars (NSs) behind (in some cases black holes (BHs) may be formed instead) \cite{Baade_1934,Colgate_1966,Smartt_2009,Janka_2017,Mezzacappa_2020,Burrows_2020,Yamada_2024,Janka_2025}. The CCSN explosion mechanism has been a topic of debate for the past decades due to its intricacies. Thanks to the improvements in the computational resources and numerical methods, a consensus is building around the neutrino-heating mechanism although there are some issues to be settled yet \cite{Muller_2017,O_Connor_2018,Kuroda_2018,Glas_2019,Müller_2020,Burrows_2020,Mezzacappa_2020,Nagakura_2020,Bollig_2021,Yamada_2024,Burrows_2024,Janka_2025,Nakamura_2025,Skyes_2025}. 

The stage between the CCSN explosion and the emergence of a cold NS, referred to as the proto-neutron star (PNS) cooling phase, has also been attracting attention. Although it was studied independently of the supernova explosion preceding it in the past \cite{Suwa_2019,Fischer_2020,Nakazato_2020,Nagakura_2021,Alp_2022,Pascal_2022,vartanyan_2023,Akaho_2024ApJ...960..116A}, it is changing thanks to recent CCSNe simulations that are long enough to cover the early stage of the PNS cooling \cite{Muller_2019,Nagakura_2021,vartanyan_2023,sykes_2024}. Just after the CCSN explosion, the PNS is very hot and lepton-rich and a large amount of energy (\(2-5\times 10^{53}\)ergs) is radiated away as neutrinos by the time when the PNS finally becomes a cold NS \cite{Roberts_2017,Fischer_2020,Nakazato_2020,Alp_2022,vartanyan_2023}. The neutrino signals from the PNS cooling phase as well as those preceding it provide invaluable information about the supernovae physics as well as the properties of the dense and hot nuclear matter and progenitor stars \cite{Raffelt_1996,Furusawa_2023}. 

The first and so far unique detection of neutrinos from a CCSN explosion is the one of those from SN 1987A, which occurred in the Large Magellanic Cloud  \cite{Hirata_1987,Haines_1988}. Although the nature of the remnant is still not clear \cite{Fransson_2024}, since there is no direct observation of a neutron star, it is rather certain that a PNS was initially formed. Most of the detected neutrinos are thought to be emitted during the PNS cooling phase \cite{Roberts_2017}. 

It is known that massive stars are rotating rather rapidly on the main sequence in general and are suggested to have a bimodal distribution of angular velocity \cite{Guthrie_1982,Attridge_1992,Choi_1996,Herbst_2001,Barnes_2003,Royer_2007,Maeder_2009,Zorec_2012,D'Antona_2015,Bastian_2020,Kamann_2021}. The faster component may be originated from the merger of binaries \cite{Kamann_2021}. Although the transfer and loss of angular momentum in their subsequent evolutions is a subject of intensive research \cite{Maeder_2009,Mink_2013,Leung_2021,Wu_2022,henneco_2023,marchant_2023}, the rotation rate of the massive star core prior to its collapse is still very uncertain \cite{Gormaz_Matamala_2023}. If they are still rotating rapidly, then the PNS produced by their collapse is likely to rotate fast also \cite{Postnov_2016,Gilkis_2017,Fujibayashi_2021,Umeda_2024,You_2024}. In that case, the neutrino emission in the subsequent cooling phase should be non-spherical, since rapidly-rotating stars are flattened by centrifugal forces. In this paper we would like to quantify this. Although they are not a direct target of this paper, rapidly-rotating hot neutron stars are thought to be formed in the NS merger \cite{Shibata_2019,Zhang_2019,Margalit_2022}. They will also emit neutrinos copiously and cool. We expect that asymmetry in the neutrino emission will be qualitatively similar to the asymmetry considered in this paper for the rapidly-rotating PNS.

The cooling phase of PNS is dictated by the neutrino transfer inside and it lasts for \(\sim 30\)s until the PNS becomes transparent for neutrinos \cite{Nakazato_2013,Geppert_2016,Nakazato_2020,Li_2021,Sumiyoshi_2023}. The PNS cooling was studied mostly in 1D under spherical symmetry in the past  \cite{Muller_2017, O_Connor_2018,Fischer_2020,Nakazato_2020,Li_2021,Sumiyoshi_2023,Nakazato_2024ApJ...975...71N} and only a small number of attempts have been made to study the rotational case. Although they employed some approximations and simplifications, such as slow and/or uniform-rotation \cite{Goussard_1999,Sumiyoshi_1999,Strobel_1999,Ye_2006,Villain_2004,Camelio_2016,Wang_2022,Salinas_2024}, these studies found that rotation has an impact on the PNS structure and stability \cite{Buellet_2023}. The total neutrino luminosity, on the other hand, was not much affected. As mentioned before, recent long-term CCSN simulations cover the earliest stages of the PNS cooling (\( \lesssim 2\)s) \cite{Nagakura_2018,Harada_2019,Harada_2020,Nagakura_2021b,Nagakura_2021,Choi_2025}. Although they employed a sophisticated multi-dimensional treatment of neutrino transfer, they considered mostly slow rotators.

In this paper, we study the anisotropic emissions of neutrinos from rapidly-rotating PNS's in their cooling phase. Since our primary interest is their global asymmetry, we ignore the matter motion and compute neutrino transfer alone on top of the fixed matter background until the neutrino distributions become steady. This may be justified, since the time for the neutrino distributions to reach steady states, \(\ \ll 1\) s, is shorter than the cooling timescale. We employ a full general relativistic, multi-dimensional Boltzmann-radiation-hydrodynamics solver based on the discrete ordinate method that we have been developing over the years for supernova simulations \cite{Sumiyoshi_2012,Nagakura_2014,Nagakura_2017,Akaho_2021} although its hydrodynamics capability is not used in this paper (but see \cite{Akaho_2023,Akaho_2025arXiv250607017A} for general relativistic radiation-hydro dynamic simulations). We perform two-dimensional simulations in axisymmetry for the rotational configurations of PNS that we build numerically by adding either a uniform- or a differential-rotation by hand to spherically symmetric matter profiles extracted from a 1D PNS cooling model \cite{Sugiura_2022} at \(t=2, 6, 10, 20\)  and \(30\)s after a supernova explosion. For reference, we also calculated non-rotation models for the original matter profiles taken from the PNS cooling calculations \cite{Sugiura_2022,Fujisawa_2025}. General relativity is fully taken into account, with the metrics given as a part of the background information. By ignoring the matter motion, we cannot address such issues as convection in PNS, which we expect to occur indeed. Note that the criterion for convection is also modified by rotation \cite{Tassoul_1979} and the convective pattern will depend on the latitude. Hence the anisotropic neutrino emissions studied in this paper will certainly be changed in the presence of convection. This paper is the first of a series, though, and we will address this and other issues one by one later.

This paper is organized as follows. In Section~\ref{NumMeth} the numerical methods employed are described and the models are summarized. In Section~\ref{Res} the results obtained from the simulations are presented and detailed analyses are given to the neutrino luminosity and the energy spectra based on the neutrino-hemisphere that will be defined later; the emission of deci-Hz gravitational waves is calculated and the possibility of neutrino fast flavor conversions is also touched. Finally, in Section~\ref{Sum} conclusions are given with some discussions.

Unless otherwise stated, we work in units with \(c=G=\hbar=1\) with \(c\), \(G\), and \(\hbar\) being the light speed, the gravitational constant, and the reduced Planck constant respectively. Throughout the paper the \(-\)  \(+\) \(+\) \(+\) metric signature is used; Greek (\(\alpha, \beta, \mu, \nu\)) and Latin (\(i, j, k\)) letters are used to denote indices that run from 0-3 and 1-3, respectively. In addition, neutrinos are assumed to be massless as their minuscule masses play no role\footnote{The neutrino masses are not essential even in the neutrino oscillations as long as the fast flavor conversion is concerned \cite{Dasgupta_2022}.}.

\section{Numerical Methods and Models}\label{NumMeth}
The numerical construction of rotational models is explained in Section \ref{NumMeth-rot}; the details of the Boltzmann solver are described in Section \ref{NumMeth-GRBoltz}. In Sections \ref{NumMeth-Nulum}, \ref{NumMeth-GW} and \ref{NumMeth-FFI} the formulation and method of calculations of the neutrino luminosity and neutrino-hemisphere position, gravitational waves and fast flavor instabilities, respectively, are described.

\subsection{Numerical construction of rotational models} \label{NumMeth-rot}

In this paper, we prepare models of rapidly-rotating PNS by ourselves, since there are no available ones obtained from CCSN simulations. We employ a general relativistic version of the versatile code developed rather recently by Fujisawa \cite{Fujisawa_2015,Fujisawa_2025}. It can numerically build axisymmetric, baroclinic rotating stars in general relativity. It solves alternately the gravitational field and matter profile until full convergence is reached. Note that the neutrino contribution is neglected in constructing the PNS configuration. Below we give a concise explanation of the method.

The line element of a stationary, axisymmetric spacetime of a rotating star without meridional motions is expressed in general as \cite{Butterworth_1976}
\begin{align}
ds^2=&-e^{\Gamma+\xi}dt^2+e^{2\psi}(dr^2+d\theta^2) \notag\\
     &+e^{\Gamma-\xi}r^2 \sin^2\theta(d\phi-\omega dt)^2,
\end{align}

\noindent where the metric components \(\Gamma\), \(\xi\), \(\psi\) and \(\omega\) are functions of \(r\) and \(\theta\).

The Euler equations are written as

\begin{equation} \label{eq:Euler1}
(\epsilon+p)^{-1} \frac{\partial p}{\partial r} + \frac{\partial}{\partial r} \left( \frac{ \Gamma + \xi }{2}\right) -\frac{v}{1-v^2}\frac{\partial v}{ \partial r} + F\frac{\partial \Omega}{\partial r} =0,
\end{equation}

\begin{equation} \label{eq:Euler2}
(\epsilon+p)^{-1} \frac{1}{r}\frac{\partial p}{\partial \theta} + \frac{1}{r}\frac{\partial}{\partial \theta} \left( \frac{ \Gamma + \xi }{2}\right) -\frac{v}{1-v^2}\frac{1}{r}\frac{\partial v}{ \partial \theta} + F\frac{1}{r}\frac{\partial \Omega}{\partial \theta} =0,
\end{equation}

\noindent where \(\epsilon\) is the energy density, \(p\) is the pressure, \(\Omega\) is the angular velocity, \(v\) is the proper velocity with respect to the zero angular momentum observer (defined as \(v=(\Omega-\omega)r\sin\theta e^{-\xi}\)) and \(F\) is the specific angular momentum defined as \(F=u^t u_{\phi}\).

The Einstein equations for the metric components are cast into the following form

\begin{equation} \label{eq:Eins1}
\nabla^2 [\xi e^{\Gamma/2}]=S_{\xi}(r,\mu),
\end{equation}

\begin{equation}\label{eq:Eins2}
\left( \nabla^2 + \frac{1}{r}\partial_r-\frac{\mu}{r^2}\partial_{\mu}\right)[\Gamma e^{\Gamma/2}] = S_{\Gamma}(r,\mu),
\end{equation}

\begin{equation}\label{eq:Eins3}
\left( \nabla^2 + \frac{2}{r}\partial_r-\frac{2\mu}{r^2}\partial_{\mu}\right)[\omega e^{(\Gamma-2\xi)/2}] = S_{\omega}(r,\mu),
\end{equation}

\begin{equation}\label{eq:Eins4}
\alpha_{,\mu} = S_{\alpha}(r,\mu),
\end{equation}

\noindent where \(\nabla^2\) is the Laplacian in the flat 3D space and \(\mu=\cos \theta\). \(S_\xi\), \(S_\Gamma\), \(S_\omega\) and \(S_\alpha\) are the source terms given, respectively, by

\begin{align}
S_\xi & = e^{\Gamma/2} \Biggl\{ 8\pi e^{2\psi} (\epsilon+p)\frac{1+v^2}{1-v^2} \notag\\
&+r^2(1-\mu^2)e^{-2\xi} \left( \omega^2_{,r}+\frac{1-\mu^2}{2}\omega^2_{,\mu}\right) \notag\\ 
&+\frac{1}{r}\Gamma_{,r}-\frac{\mu}{r^2}\Gamma{,\mu} +\frac{\xi}{2}\Biggl[ 16\pi e^{2\psi}p \notag\\
&- \Gamma_{,r} \left( \frac{\Gamma_{,r}}{2}+\frac{1}{r}\right) -\frac{1}{r^2}\Gamma_{,\mu} \left( \frac{1-\mu^2}{r^2}\Gamma_{,\mu}-\mu \right) \Biggr] \Biggr\},
\end{align}
\begin{align}
S_\Gamma(r,\mu) & = e^{\Gamma/2} \Biggl\{ 16\pi e^{2\xi}p  \notag\\
&- \frac{\Gamma}{2}\Biggl[ 16\pi e^{2\xi}p - \frac{1-\mu^2}{2r^2}\Gamma^2_{,\mu} \Biggr] \Biggr\},
\end{align}
\begin{align}
S_\omega(r,\mu) & =e^{(\Gamma-2\xi)/2} \Biggl\{ -16\pi e^{2\xi} \frac{(\Omega-\omega)(\epsilon+p)}{1-v^2} \notag\\
&+\omega \Biggl[ -8\pi e^{2\xi} \left( \frac{(\epsilon+p)(1+v^2)}{1-v^2}-p \right) \notag\\
& -\frac{1}{r} \left( 2\xi_{,r} + \frac{1}{2}\Gamma_{,r} \right) + \frac{\mu}{r^2} \left( 2\xi_{,\mu} + \frac{1}{2}\Gamma_{,\mu} \right) \notag\\
&+\frac{1}{4} (4\xi^2_{,r}-\Gamma^2_{,r})+\frac{1-\mu^2}{4r^2}(4\xi^2_{,\mu}-\Gamma^2_{,\mu}) \notag\\
& -r^2(1-\mu^2)e^{-2\xi} \left( \omega^2_{,r} + \frac{1-\mu^2}{r^2}\omega^2_{,\mu} \right) \Biggr] \Biggr\},
\end{align}
\begin{align}
S_{\alpha}(r,\mu) &=-\frac{1}{2}(\xi_{,\mu}+\Gamma_{,\mu})-\{(1-\mu^2)(1+r\Gamma_{,r})^2 \notag\\
& + [\mu-(1-\mu^2)^2\Gamma_{,\mu}]^2\}^{-1} \notag\\
& \Biggl[ \frac{1}{2}[r^2(\Gamma_{,rr}+\Gamma^2_{,r})-(1-\mu^2)(\Gamma_{,\mu\mu}+\Gamma^2_{,\mu})] \notag\\
&[-\mu+(1-\mu^2)\Gamma_{,\mu}]+r\Gamma_{,r} \Bigl[ \frac{1}{2}\mu + \mu r \Gamma_{,r} \notag\\
& + \frac{1}{2}(1-\mu^2)\Gamma_{,\mu}  \Bigr] + \frac{3}{2}\Gamma_{,\mu}[-\mu^2+\mu(1-\mu^2)\Gamma_{,\mu}] \notag\\
& -r(1-\mu^2)(\Gamma_{,r\mu}+\Gamma_{,r}\Gamma_{,\mu})(1+r\Gamma_{,r}) \notag\\
& -\frac{\mu r^2}{4} (\xi_{,r}+\Gamma_{,r})^2-\frac{r}{2}(1-\mu^2)(\xi_{,r}+\Gamma_{,r})(\xi_{,\mu}+\Gamma_{,\mu}) \notag\\
& +\frac{\mu}{4}(1-\mu^2)(\xi_{,\mu}+\Gamma_{,\mu})^2 \notag\\
&-\frac{r^2}{2}(1-\mu^2)\Gamma_{,r}(\xi_{,r}+\Gamma_{,r})(\xi_{,\mu}+\Gamma_{,\mu}) \notag\\
&+\frac{\Gamma_{,\mu}}{4}(1-\mu^2)[r^2(\xi_{,r}+\gamma_{,r})^2-(1-\mu^2)(\xi_{,\mu}+\gamma_{,\mu})^2]\notag\\
& + (1-\mu^2)e^{-2\xi} \Bigl\{ \frac{1}{4}r^4\mu\omega^2_{,r}  +\frac{r^3}{2}(1-\mu^2)\omega_{,r}\omega_{,\mu} \notag\\
& -\frac{r^2\mu}{4}(1-\mu^2)\omega^2_{,\mu} + \frac{r^4}{4}(1-\mu^2)\Gamma_{,r}\omega_{,r}\omega_{,\mu} \notag\\
& -\frac{r^2}{4}(1-\mu^2)\Gamma_{,\mu}[r^2\omega^2_{,r}-(1-\mu^2)\omega^2_{,\mu}] \Bigr\}\Biggr].
\end{align}

In order to impose the boundary condition at spatial infinity properly, we use a compactified  radial coordinate \(s\) defined by
\begin{equation}
r = r_e \left( \frac{s}{1-s} \right),
\end{equation}

\noindent where the \(r_e\) is the equatorial radius. The original radial coordinate \(r\) covering \( [0, \infty]\) is transformed to \(s\) in the range of \([0, 1]\) in the compactified coordinate. We deploy \(257\) radial grid points in the range of \([0,\ r_e]\) in \(r\) or $[0,\ 1/2]$ in \(s\) and \(257\) points in the \([r_e,\ \infty]\) range in \(r\) or \([1/2,\ 1]\) in \(s\). Since the PNS is a bit more bloated than the cold NS because the former is still hot and lepton-rich. The density gradient is less steep in the former than in the latter although we deploy many grid points near the PNS surface. In the \(\mu\) direction, we use a uniform grid with \(129\) points.

As mentioned earlier, the equations for the gravitational field (Eq. \ref{eq:Eins1} through \ref{eq:Eins4}) and those for the matter profile are alternately solved. For the latter, we employ the \(\phi\)-component of the curl of the equations of motion:

\begin{align} \label{eq:euler-ph}
& (\epsilon+p)^2 \left( \frac{\partial (\epsilon+p)}{\partial \theta}\frac{\partial p}{\partial r} - \frac{\partial (\epsilon+p)}{\partial r}\frac{\partial p}{\partial \theta} 
 \right) \notag\\
& + \frac{\partial F}{\partial \theta} \frac{\partial \Omega}{\partial r} - \frac{\partial F}{\partial r} \frac{\partial \Omega}{\partial \theta} = 0
\end{align}

\noindent instead of Eq. \ref{eq:Euler2}.

\begin{table*}
    \centering
    \begin{tabular}{|c|c|c|c|c|c|c|c|}
         \hline
         Model & \(\rho_c\) (g/cm$^3$) & \(M_{G} (M_{\odot})\) & \(M_b (M_{\odot})\) & \(P_e\) (ms) & \(J\) (g cm$^2$/s) & \(|T/W|\) & \(r_p/r_{eq}\) \\
         \hline
         NR02 & \(6.10\times 10^{14}\) & 1.430 & 1.47& 0 & 0 & 0 & 1.00 \\
         NR06 & \(6.75\times 10^{14}\) & 1.400 & 1.47& 0 & 0 & 0 & 1.00 \\
         NR10 & \(6.96\times 10^{14}\) & 1.386 & 1.47& 0 & 0 & 0 & 1.00 \\
         NR20 & \(7.05\times 10^{14}\) & 1.369 & 1.47& 0 & 0 & 0 & 1.00 \\
         NR30 & \(7.18\times 10^{14}\) & 1.360 & 1.47& 0 & 0 & 0 & 1.00 \\
         \hline
         UR02 & \(4.19\times 10^{14}\) & 1.430 & 1.47& 4.11 &  \(7.74\times 10^{48}\) & \(3.75\times 10^{-2}\) & 0.70\\
         UR06 & \(5.80\times 10^{14}\) & 1.400 & 1.47& 2.75 &  \(7.68\times 10^{48}\) & \(4.47\times 10^{-2}\) & 0.75\\
         UR10 & \(6.08\times 10^{14}\) & 1.387 & 1.47& 2.43 & \(7.75\times 10^{48}\) & \(4.85\times 10^{-2}\) & 0.75\\
         UR20 & \(6.44\times 10^{14}\) & 1.369 & 1.47& 2.14 & \(7.75\times 10^{48}\) & \(5.21\times 10^{-2}\) & 0.78\\
         UR30 & \(6.75\times 10^{14}\) & 1.360 & 1.47& 2.03 & \(7.76\times 10^{48}\) & \(5.42\times 10^{-2}\) & 0.76\\
         \hline
         DR02 & \(4.07\times 10^{14}\) & 1.430 & 1.47& 2.47 & \(7.65\times 10^{48}\) & \(3.71\times 10^{-2}\) & 0.82\\
         DR06 & \(5.69\times 10^{14}\) & 1.400 & 1.47& 1.71 & \(7.69\times 10^{48}\) & \(4.54\times 10^{-2}\) & 0.82\\
         DR10 & \(5.98\times 10^{14}\) & 1.386 & 1.47& 1.52 & \(7.76\times 10^{48}\) & \(4.75\times 10^{-2}\) & 0.82\\
         DR20 & \(6.37\times 10^{14}\) & 1.369 & 1.47& 1.36 & \(7.78\times 10^{48}\) & \(5.33\times 10^{-2}\) & 0.81\\
         DR30 & \(6.70\times 10^{14}\) & 1.360 & 1.47& 1.30 & \(7.74\times 10^{48}\) & \(5.50\times 10^{-2}\) & 0.80\\
         \hline
    \end{tabular}
    \caption{Characteristic quantities of all models. The model names are given such a way that the first two letters indicate either no- (NR) or uniform- (UR) or differential-rotation (DR); the last two numbers denote the time after explosion of the snapshot. The characteristic quantities shown are the central density \(\rho_c\), the gravitational mass \(M_G\), the baryonic mass \(M_b\), the rotation period at the equatorial surface \(P_e\), the total angular momentum \(J\), the ratio of rotation energy to gravitational mass energy \(|T/W|\) and the ratio of the polar radius to the equatorial radius \(r_p/r_{eq}\).}
    \label{tab:carquan}
\end{table*}

The concrete procedure is the following. We start with an arbitrarily chosen configuration, normally a non-rotating one as a solution of the Tolman–Oppenheimer–Volkoff (TOV) equation. We solve equations \ref{eq:Eins1} through \ref{eq:Eins4} to obtain the metric components for this initial configuration as a very crude approximation. We set an angular velocity profile on the equator \(\Omega=\Omega(\varpi)\) with \(\varpi\) being the cylindrical radius and integrate Eq. \ref{eq:euler-ph} in the \(\theta\) direction to obtain \(\Omega(r, \theta)\). Then we integrate Eq. \ref{eq:Euler1} in the radial direction along each radial ray to obtain the matter distribution. For this new configuration, we calculate the metric components anew and repeat the procedure until convergence is obtained. 
\begin{figure}
    \centering
    \includegraphics[width=0.8\linewidth]{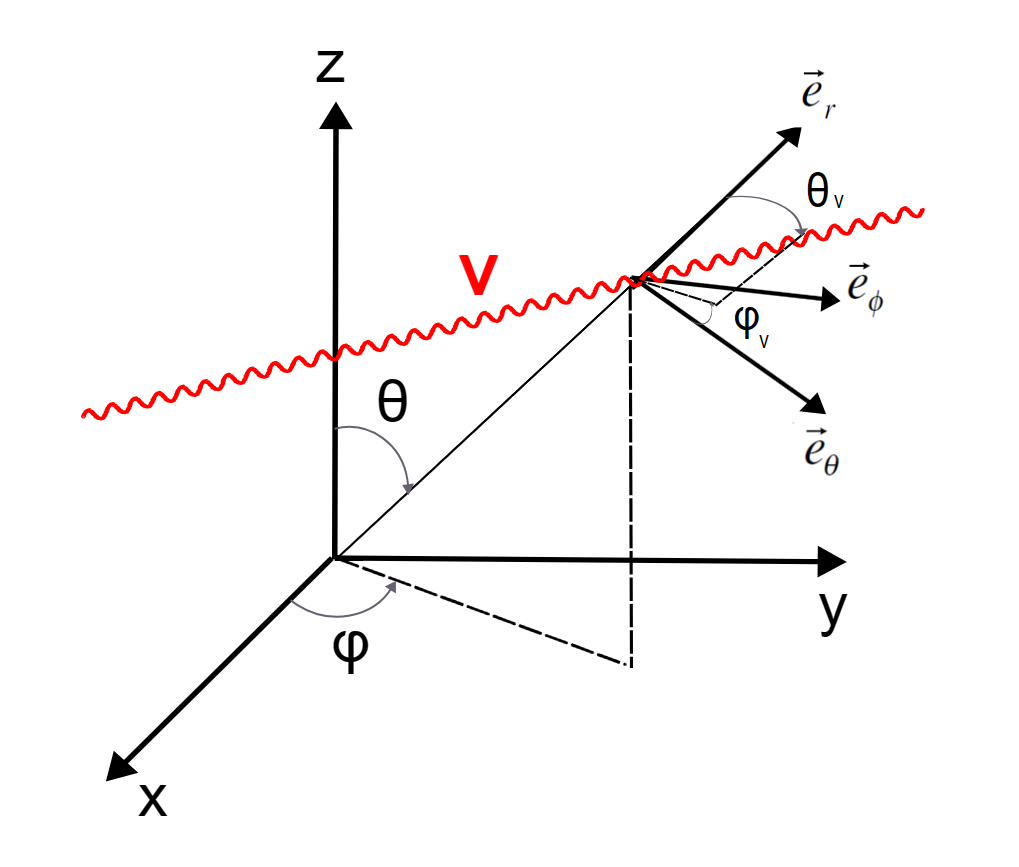}
    \caption{Schematic picture of the coordinate system employed in this paper. The momentum space coordinates are locally defined with respect to the tetrad deployed there.}
    \label{fig:axis}
\end{figure}
In the general baroclinic case, we employ that the entropy distribution in space is given. Note that no first integral is required in this formulation unlike most of other methods \cite{Fujisawa_2015}. In this paper, we assume pseudo barotropic EOS's that differ from the EOS employed in the PNS cooling simulations or in the neutrino transfer calculations and are constructed as follows: the density \(\rho(r)\), pressure \(p(r)\), specific entropy \(s(r)\) and electron fraction \(Y_e(r)\) are read off from the non-rotating PNS-cooling models of \cite{Nakazato_2018} at the designated post-explosion times (\(t=2, 6, 10, 20, 30\)s) and regarded the correspondences \(\rho(r)\) vs \(p(r)\), \(s(r)\) and \(Y_e (r)\) at the same radial point as the pseudo barotropic EOS. We think that this method reproduces better the original features in the profiles. Then the rotational equilibria is solved for the given rotation laws. The resultant rotations are cylindrical (modulo general relativistic corrections). In this paper, we consider a model sequence of uniform-rotation, i.e., \(\Omega(\varpi)=\) const., and another sequence of differential-rotation given by \(\Omega(\varpi)=\frac{\Omega_c}{1+\varpi^2/r_e^2}\) with \(r_e\) being the equatorial radius, in addition to a non-rotational model sequence for comparison. Along each sequence, the total angular momentum (and the baryonic mass) is approximately constant as mass accretion or loss as well as magnetic fields are neglected in this paper. The values of angular momentum adopted in this paper are corresponding to the ratio of rotation energy to gravitational energy, \(|T/W|\), of \(\sim0.05\). The baryonic mass is \(M_b\sim 1.47M_{\odot}\). The characteristic quantities (the central density \(\rho_c\), the gravitational mass \(M_G\), the rotation period at the equatorial plane \(P_e\), the total angular momentum \(J\), the ratio of rotation energy to gravitational mass energy \(|T/W|\) and the ratio of the polar radius to the equatorial radius \(r_p/r_{eq}\)) in addition to the baryonic mass \(M_b\) are summarized for all models in Table \ref{tab:carquan}.

\subsection{Neutrino Transfer with a General Relativistic  Boltzmann Equation Solver}\label{NumMeth-GRBoltz}

For the radiative transport of neutrinos, we solve numerically the general relativistic Boltzmann equation cast in its conservative form \cite{Shibata_2013}:

\begin{align}
&\frac{1}{\sqrt{-g}}\frac{\partial}{\partial x^\mu}\Bigg|_{q_{i}}\left[\left(e^\mu_{(0)}+\sum_{i=1}^{3}l_ie^\mu_{(i)}\right)\sqrt{-g}f\right] \notag\\ 
&-\frac{1}{\epsilon^2}\frac{\partial}{\partial\epsilon}\left(\epsilon^3f\omega_{(0)}\right)+\frac{1}{\sin{\theta_{\nu}}}\frac{\partial}{\partial \theta_{\nu}}\left(\sin{\theta_{\nu}}f\omega_{(\theta_{\nu})}\right) \notag\\
&-\frac{1}{\sin^2{\theta_{\nu}}}\frac{\partial}{\partial \phi_{\nu}}\left(f\omega_{(\phi_{\nu})}\right)=S_{rad},
\end{align}

\noindent where \(x^{\mu}\) are the spacetime coordinates;  \(q_i\) represents the momentum space coordinates: \(q_1=\epsilon\), \(q_2=\theta_{\nu}\), \(q_3=\phi_{\nu}\) with \(\theta_{\nu}\) and \(\phi_{\nu}\) being the zenith and azimuth angles, respectively, in momentum space (see Fig. \ref{fig:axis}); \(\epsilon\) is the neutrino energy defined as \(\epsilon=-p_{\mu}e^{\mu}_{(0)}\), with \(p^{\mu}\) the neutrino four momentum and \(e_{(\alpha)}^{\mu}\) being the local tetrad given below; \(f\) is the neutrino distribution function; \(g\) is the determinant of the spacetime metric tensor; and \(S_{rad}\) is the collision term. The set of the tetrad basis \(e^{\mu}_{(\alpha)}\) (\(\alpha=0,1,2,3\)) is given as follows
\begin{equation}\label{eq:tetrad1}
e^{\mu}_{(0)}=n^{\mu},
\end{equation}

\begin{equation}\label{eq:tetrad2}
e^{\mu}_{(1)}=\gamma^{-\frac{1}{2}}_{rr}\left(\frac{\partial}{\partial r}\right)^{\mu},
\end{equation}

\begin{align}\label{eq:tetrad3}
  e^{\mu}_{(2)}=&-\frac{\gamma_{r\theta}}{\sqrt{\gamma_{rr}(\gamma_{rr}\gamma_{\theta\theta}-\gamma_{r\theta}^2)}}\left(\frac{\partial}{\partial r}\right)^{\mu} \notag\\
  &+\sqrt{\frac{\gamma_{rr}}{\gamma_{rr}\gamma_{\theta\theta}-\gamma_{r\theta}^2}}\left(\frac{\partial}{\partial \theta}\right)^{\mu}
\end{align}

\begin{align}\label{eq:tetrad4}
    e^{\mu}_{(3)}&=\frac{\gamma^{r\phi}}{\sqrt{\gamma^{\phi\phi}}}\left(\frac{\partial}{\partial r}\right)^{\mu}+\frac{\gamma^{\theta\phi}}{\sqrt{\gamma^{\phi\phi}}}\left(\frac{\partial}{\partial \theta}\right)^{\mu} \notag\\
    & +\sqrt{\gamma^{\phi\phi}}\left(\frac{\partial}{\partial \phi}\right)^{\mu},
\end{align}

\noindent where \(\gamma_{ij}\) is the spatial component of the metric on the polar coordinates and \(n^{\nu}\) is the timelike unit vector normal to the time-constant hypersurface. The directional cosines in momentum space \(l_{(i)}\) are defined as
\begin{equation}
l_{(1)}=\cos{\theta_{\nu}},
\end{equation}
\begin{equation}
l_{(2)}=\sin{\theta_{\nu}}\cos{\phi_{\nu}},
\end{equation}
\begin{equation}
l_{(3)}=\sin{\theta_{\nu}}\sin{\phi_{\nu}}.
\end{equation}

\noindent Finally the factors \(\omega_{i}\) are defined as
\begin{equation}
\omega_{i}=\epsilon^{-2}p^{\mu}p_{\nu}\nabla_{\mu}e^{\mu}_{(i)}.
\end{equation}

\noindent These equations are finite-differenced in all coordinate directions. For the detail we refer to \cite{Sumiyoshi_2012,Nagakura_2014,Nagakura_2017,Akaho_2021}.

The original matter profiles and metric tensor components are extended up to \(40\)km in the simulations to avoid a discontinuous transition to vacuum. The metric tensor is smoothly extrapolated to the flat metric. The matter quantities, on the other hand, are assumed to be constant in this extended region except for the density, which is decreased exponentially down to the floor value, \(1\times10^{-7}\)g/cm$^3$. This is done for numerical reasons. Since this hydrodynamics are not solved in this study and this extension is done after the rotational equilibrium is obtained, it has no influence on the matter profiles in the PNS. We also hope that the extension approximates, albeit very crudely, the real situation surrounding the PNS. On top of the fixed matter profile thus obtained, the time evolution of the neutrino distribution function is followed until a steady state is reached. Note that not only the density but the specific entropy \(s\) and electron fraction $Y_e$ are also fixed during the neutrino transfer calculations.

We deploy the non-uniform radial mesh with \(N_r=128\) grid points that covers the range \(r\in [0,40]\)km. The grid points are concentrated near the PNS surface, where the scale height is very short even in the PNS. The number of spatial angular mesh points is \(N_{\theta}=32\). Due to the axisymmetry, there is no dependence on \(\phi\). As for the neutrino angles, the number of grid points is \(N_{\theta_{\nu}}=10\) and \(N_{\phi_{\nu}}=6\). The number of energy grid points is \(N_{\epsilon}=46\) and covers the range \(\epsilon \in [0,250]\)MeV.

Three neutrino species are taken into account: \(\nu_e\), \(\bar{\nu}_e\) and \(\nu_x\), with the last one denoting collectively the heavy-lepton neutrinos (\(\tau\) and \(\mu\) neutrinos and their anti-particles). As for the neutrino reactions, we employ the same set as in  \cite{Sumiyoshi_2012,Akaho_2021}, i.e., those essentially based on Bruenn's rates \cite{Bruenn_1985,Sumiyoshi_2005}. In evaluating these rates, we use the nuclear equation of state (EOS) by Togashi, the same EOS employed in the non-rotational PNS cooling  model \cite{Togashi_2014,Togashi_2017}.

\subsection{Observer-Dependent Neutrino Luminosity and Neutrino-Hemisphere}\label{NumMeth-Nulum}

The local number and flux densities of neutrinos are given, respectively, as

\begin{equation} \label{eq-numdens}
n_{\nu} = \frac{1}{(2 \pi)^3} \int \epsilon^2 d\epsilon \int d\Omega_\nu f,
\end{equation}

\begin{equation} \label{eq-numbflux}
F_{\nu} = \frac{1}{(2 \pi)^3} \int \epsilon^2 d\epsilon \int d\Omega_\nu f \cos\theta_{\nu}.
\end{equation}

\noindent In this paper we are interested in the anisotropy of neutrino radiations from rapidly-rotating PNS's. In order to quantify it and use the results later for the evaluation of gravitational waves, we consider the observer-dependent luminosity defined as follows:
\begin{align} \label{eq:obs-dependentLum}
\frac{dL_\nu(\mu_o)}{d\Omega_o}=&\int R^2 \alpha(R,\Omega) \,d\Omega 
\int \frac{\epsilon^3 }{(2\pi\hbar c)^3} \,  d\epsilon  \notag\\ 
& f(R,\Omega,\Omega_{\nu}(\Omega,\Omega_o),\epsilon)c\mu_{\nu}(\Omega,\Omega_o),
\end{align}

\noindent where  \(R\) is the radius of the sphere outside PNS, at which this luminosity is evaluated and \(\Omega\) denotes the angular position on it; \(\Omega_o\) refers to the direction of the observer at a large distance; \(\Omega_\nu\) is the direction of neutrino 3-momentum (\(\mu_\nu=\cos\theta\)) and \(\epsilon\) is the neutrino energy; \(\alpha=\sqrt{g_{00}+g_{03}^2/g_{33}}\) is the lapse function that takes account of the redshift correction. Note that \(R\) is arbitrary as long as the sphere is located in the region, where neutrino-matter interactions are effectively turned off and neutrinos are streaming freely. We choose \(R=40\)km. If we set it to a larger value, we would waste numerical resources and, moreover, highly forward-peaked and narrow angular distributions in momentum space at such large distances could not be resolved with the current numbers of the grid points (10 \((\theta_{\nu}) \times 6 (\phi_{\nu}))\). Without a loss of generality, we can assume that the observer is located in the \(x-z\) plane (\(\phi_o=0)\). The integration with respect to \(\Omega\) is done over the hemisphere facing the observer. From the definition of the observer-dependent luminosity, the total neutrino luminosity emitted by the PNS is given as
\begin{equation} \label{eq:totlum}
L_{\nu}=\int \frac{dL_\nu(\mu_o)}{d\Omega_o} \,d\Omega_o.
\end{equation}

\begin{figure}[t]
    \centering
    \includegraphics[width=0.85\linewidth]{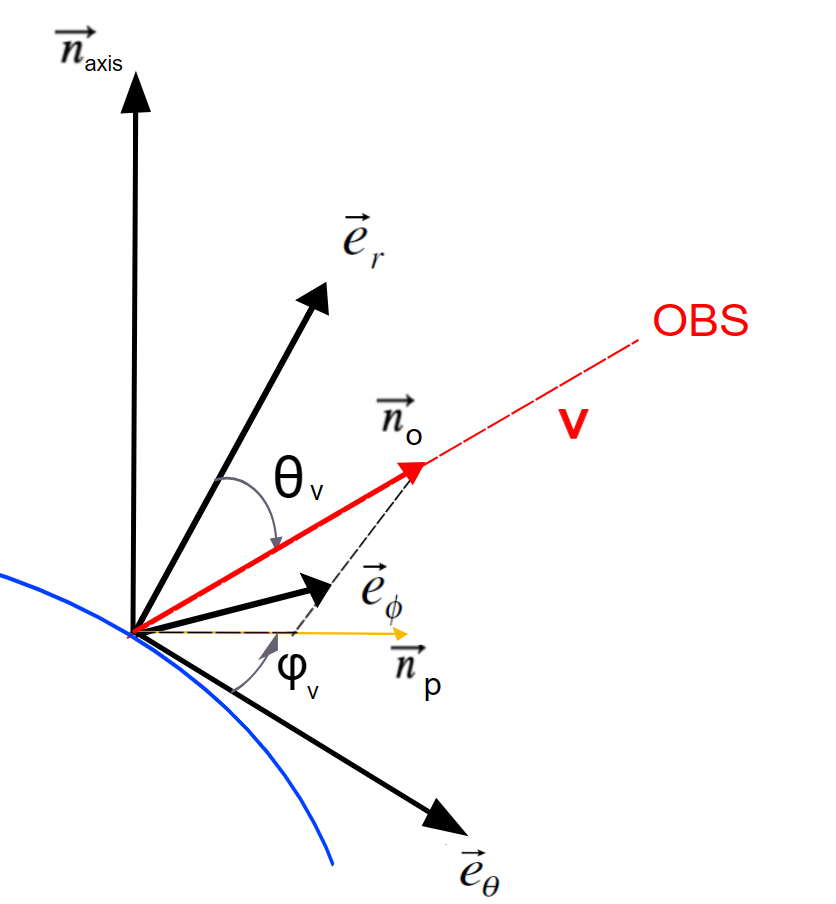}
    \caption{Schematic picture showing the relation of the observer's position and the neutrino angles: \(\vec{e}_r\), \(\vec{e}_\theta\), \(\vec{e}_\phi\) are local orthogonal unit vectors; \(\vec{n}_o\) is the unit vector in the direction of the observer; \(\vec{n}_p\) is the unit vector of the projection of \(\vec{n}_o\) in the  \(\vec{e}_\theta\)-\(\vec{e}_\phi\) plane and \(\vec{n}_{axis}\) denotes the direction of the rotation axis.}
    \label{fig:OBS}
\end{figure}

As stated before, the neutrino distribution function is a function of the angles of neutrino 3-momentum that are defined with respect to the tetrad specified locally (Eqs. \ref{eq:tetrad1}-\ref{eq:tetrad4} and see also Fig. \ref{fig:axis}). Hence we need to express it in terms of the observer angle \(\theta_o\). This can be done by calculating \(\cos \theta_{\nu}\) and \(\cos \phi_{\nu}\) as follows,
\begin{equation}\label{eq:thnu_obs}
    \cos \theta_\nu = \vec{e}_r\cdot \vec{n}_o,
\end{equation}
\begin{equation}\label{eq:phnu_obs}
    \cos \phi_\nu = \vec{n}_p \cdot \vec{e}_\theta,
\end{equation}
\noindent where \(\vec{e}_r\) and \(\vec{e}_\theta\) are the unit vectors pointing in the radial and azimuthal directions, respectively, \(\vec{n}_o\) is the unit vector pointing towards the observer and \(\vec{n}_p\) is the unit vector projected onto the \(\vec{e}_\theta\)-\(\vec{e}_\phi\) plane. These vectors are given as

\begin{equation}
\vec{e}_r =
\begin{pmatrix}
\sin \theta \cos \phi \\
\sin \theta \sin \phi \\
\cos{\theta}
\end{pmatrix},
\end{equation}

\begin{equation}
\vec{e}_{\theta} =
\begin{pmatrix}
\cos \theta \cos \phi \\
\cos \theta \sin \phi \\
-\sin{\theta}
\end{pmatrix},
\end{equation}

\begin{equation}
\vec{e}_{\phi} =
\begin{pmatrix}
-\sin \phi \\
\cos \phi \\
0
\end{pmatrix},
\end{equation}

\begin{equation}
\vec{n}_0 = 
\begin{pmatrix}
\sin \theta_o \\
0 \\
\cos{\theta_o}
\end{pmatrix},
\end{equation}

\begin{equation}
\vec{n}_p = \frac{\vec{n}_o-\vec{e}_r \cos \theta_{\nu}}{\sin \theta_\nu}.
\end{equation}
\begin{figure}
     \centering
     \begin{subfigure}[b]{1\columnwidth}
         \centering
         \includegraphics[width=\columnwidth]{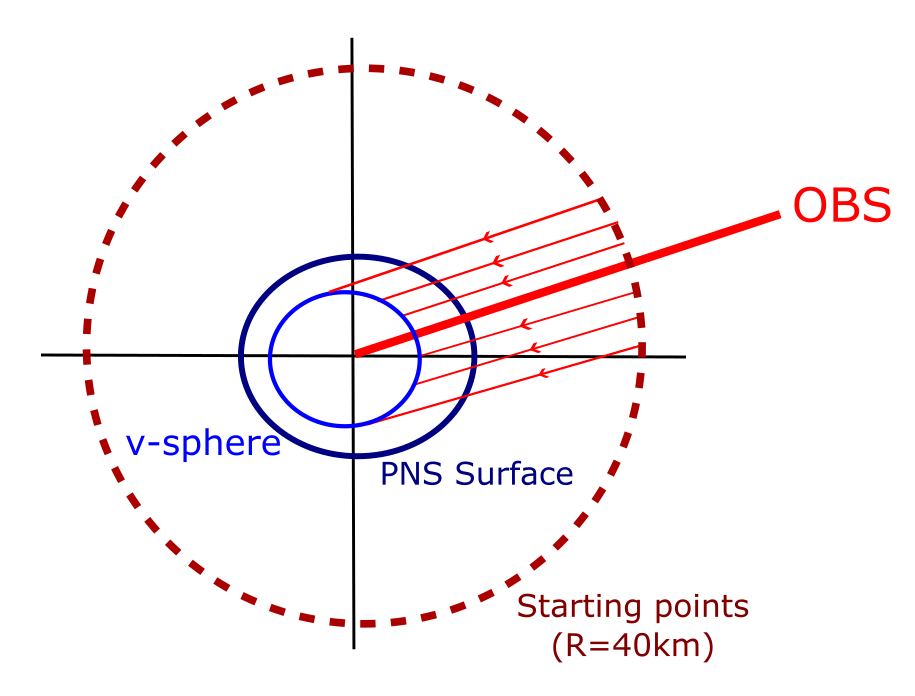}
         \caption{ }
         \label{fig:nu-sph-calc1}
     \end{subfigure}
     \centering
     \begin{subfigure}[b]{1\columnwidth}
         \centering
         \includegraphics[width=\columnwidth]{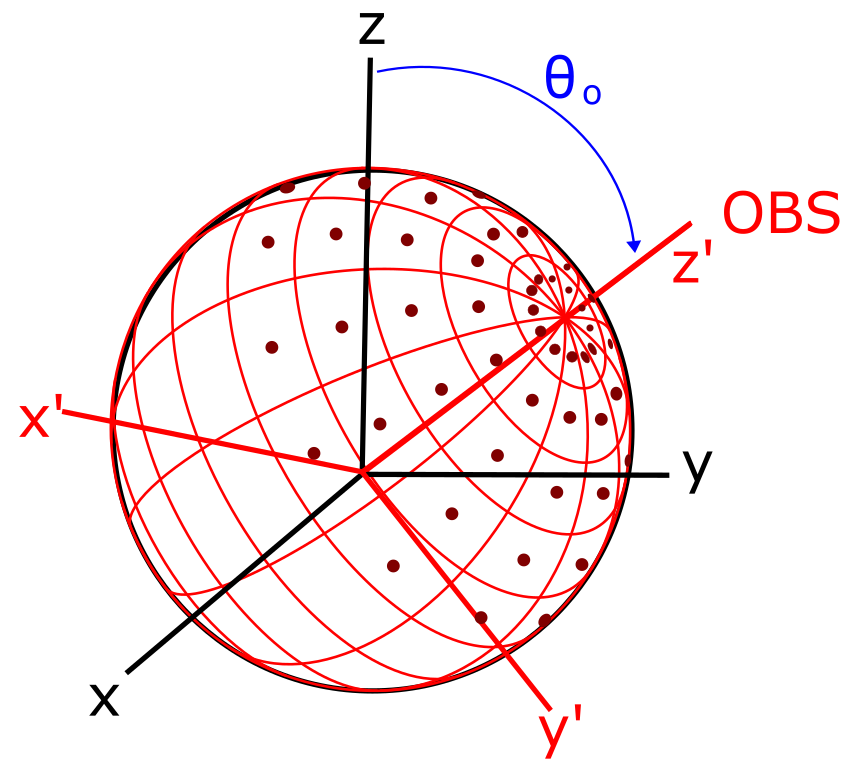}
         \caption{ }
         \label{fig:nu-sph-calc2}
     \end{subfigure}
     \caption{Scheme for locating the neutrino-hemispheric position. (a) The schematic picture of the rays, along which the optical depth is calculated. (b) The starting points of the rays (red dots).}
     \label{fig:TEST}
\end{figure}

\noindent Note that general relativity is ignored here, since it is small on the sphere, where the observer-dependent luminosity is evaluated (\(R=40\)km).

The neutrino-sphere is the spherical surface, from which neutrinos are radiated away effectively. It is normally defined as a set of points, at which the optical depth from infinity along the radial ray is equal to \(2/3\):

\begin{equation} \label{eq:tau}
\tau=-\int_\infty^r \frac{dl}{\lambda},
\end{equation}

\noindent where \(\lambda\) is the mean free path. The matter temperature as well as the local neutrino distributions on the  neutrino-sphere are useful for understanding the neutrino luminosity and spectrum. In our case, the PNS is rapidly-rotating and is not spherically symmetric but oblate and, as a result, the distributions of matter and neutrinos are also asymmetric. It is hence better to consider the neutrino-sphere observer-wise. Since the hemisphere facing the observer is important, we consider the {\it neutrino-hemisphere} for each observer direction.
It is constructed as follows. We consider again a sphere of the radius of \(40\)km, which is the outer boundary of the computational domain, and emit rays from a set of points on the sphere inward to the PNS; they are all parallel to the radial ray to the observer (see Fig. \ref{fig:nu-sph-calc1}). We then integrate the inverse mean free path as in Eq. \ref{eq:tau} but along each ray inward from its starting point on the outer boundary. Those starting points are chosen from the grid points of the spherical coordinates that are deployed on the sphere of \(r=40\)km and are rotated in such a way that the north pole is directed to the observer; only the points on the hemisphere facing the observer are chosen (Fig. \ref{fig:nu-sph-calc2}). The set of those points, at which the integral value reaches 2/3, is referred to as the observer-dependent neutrino-hemisphere in this paper. We analyze the shape of and the average neutrino energies on this hemisphere in detail.

\subsection{Gravitational Waves Emitted by Neutrinos}\label{NumMeth-GW}
It is known that anisotropic radiations of neutrinos are accompanied by emissions of gravitational waves (GWs) \cite{Epstein_1978,Turner_1978,Vartanyan_2020,Richardson_2022,Choi_2024}. The GWs emitted during the PNS cooling phase were first studied by \cite{Mukhopadhyay_2021,Fu_2022}. They showed that the GWs are emitted in the frequency range of \(0.1-10\)Hz and could have strains of \(\sim 10^{-22}-10^{-20}\) from Galactic sources.

In this paper we follow the formulation derived in \cite{Fu_2022} and evaluate those deci-Hz GWs based on our more realistic models of neutrino emissions. The waveform of the GW emitted from neutrinos radiated anisotropically is given as
\begin{equation} \label{eq-hsorig}
h_s(t,\alpha,\beta)=\frac{2G}{c^4 D} \int_0^t \,dt' \int  d\Omega_o W_S(\Omega_o,\alpha,\beta)\frac{dL_\nu(t',\Omega_o)}{d\Omega_o} , 
\end{equation}
where \(S\in(+,\times)\) specifies the polarization of the GW, \(dL_\nu/d\Omega_o\) is the observer-dependent neutrino luminosity, \(D\) is the distance to the observer from the source and \(W_S\) is the angular weight given as

\begin{equation}
W_S(\theta_o,\phi_o,\alpha,\beta)=\frac{D_S(\theta_o,\phi_o,\alpha,\beta)}{N(\theta_o,\phi_o,\alpha,\beta)},
\end{equation}

\noindent with

\begin{align}
    D_{+}&=[1+(\cos\phi_o\cos\alpha+
    \sin\phi_o\sin\alpha)\sin
    \theta_o\sin\beta\notag\\
    &+\cos\theta_o\cos\beta]([(\cos\phi_o\cos\alpha+\sin\phi_o\sin
    \alpha)\notag\\
    &\sin\theta_o\cos\beta-\cos\theta_o\sin\beta]^{2}-\sin^{2}\theta_o\notag\\
    &(\sin\phi_o\cos
    {\alpha-\cos\phi_o\sin\alpha})^{2}),
\end{align}

\begin{align}
    D_{\times}&=[1+(\cos\phi\cos\alpha+\sin\phi\sin\alpha)\sin\theta\sin\beta\notag\\    &+\cos\theta\cos\beta]2[(\cos\phi\cos\alpha+\sin\phi\sin\alpha)\\
    &\sin\theta\cos\beta\notag-\cos\theta\sin\beta]\sin\theta(\sin\phi\cos\alpha \notag\\
    &-\cos\phi\sin\alpha),
\end{align}

\begin{align}
    N&=[(\cos\phi\cos\alpha+\sin\phi\sin\alpha)\sin\theta\cos\beta\notag\\
    &-\cos\theta
    \sin\beta]^{2}+\sin^{2}\theta(\sin\phi\cos\alpha \notag\\
    &-\cos\phi\sin\alpha)^{2}.
\end{align}

\noindent In the above equations, \(\alpha \in [-\pi,\pi]\) and \(\beta \in [0,\pi]\) are the azimuth and zenith angles of the propagation direction of the GW.

The observer-dependent neutrino luminosity is expanded as
\begin{equation}
\frac{dL_\nu(t,\Omega_o)}{d\Omega_o}=\sum_{l=0}^\infty \sum_{m=-l}^{l} \frac{a_{lm}(t)}{4\pi} Y_l^m(\Omega_o),
\end{equation}
where we employ the real spherical-harmonics defined as
\begin{equation}
\boldsymbol{Y}^{m}_{l}(\theta,\phi)
=\left\{
\begin{array}{rcl}
\sqrt{2}K^{m}_{l}P^{m}_{l}(\cos\theta)\cos(m\phi) & & m \ge 0\\
\sqrt{2}K^{m}_{l}P^{m}_{l}(\cos\theta)\sin(m\phi)  & & m < 0
\end{array} \right.
\end{equation}
where \(l=0,1,2,...\) and \(m=-l,...,0,...,l\) and 
\begin{equation}
K_l^m=(-1)^{\frac{m+|m|}{2}}\sqrt{\frac{2l+1}{4\pi}\frac{(l-m)!}{(l+m)!}};
\end{equation}

\noindent \(P_l^m\)'s are the associated Legendre functions. Then Eq. \ref{eq-hsorig} becomes
\begin{equation}
h_s(t,\alpha,\beta)= \sum_{l=0}^\infty \sum_{m=-l}^{l}  h_{lm}^{amp}\Psi_{lm}^{S}(\alpha,\beta),
\end{equation}
with
\begin{equation}
h_{lm}^{amp}=\frac{2G}{c^4 R} \int_0^t \,dt' a_{lm}(t')dt',
\end{equation}
\begin{equation}
\Psi_{lm}^{S}(\alpha,\beta)=\frac{1}{4\pi}\int_{\Omega_o}\,d\Omega_o W_S(\Omega_o,\alpha,\beta)Y_l^m(\Omega_o).
\end{equation}

In this paper, the PNS is axsymmetric and there is no \(\alpha\)-dependence (and we set \(\alpha=0\) in the following) and the spherical-harmonic components with \(m \ne 0\) are all vanishing; moreover, only the \(h_{+}\) is non-vanishing. For comparison of the GW signal with the detector sensitivity,  the characteristic strain is used. This quantity denoted by \(\tilde{h}_S^c(f)\) is defined as 

\begin{equation}
\tilde{h}_S^c(f)=2f|\tilde{h}_S(f)|,
\end{equation}
where \(\tilde{h}_S(f)\) is the Fourier transform of the GW waveform \(h_s(t)\):
\begin{equation}
\tilde{h}_S(f)=\int_{-\infty}^\infty h_S(t)e^{2\pi ift}dt.
\end{equation}

\subsection{Fast Flavor Conversions of Neutrinos}\label{NumMeth-FFI}

It is well established that neutrinos have small but non-vanishing masses that are not diagonal in flavors \cite{Fukuda_1998,Gava_2008,Navas_2024} and, as a result, they change flavors as they propagate in space \cite{Gonzalez_Garcia_2003,Balantekin_2013}. When a neutrino runs through a region with copious neutrinos present, such as in PNS's, its mass is modified effectively as a refractive effect of its interactions with other neutrinos and, as a consequence, the flavor conversion is also affected. Since those other neutrinos also change their flavors, the conversion process becomes nonlinear, showing a rich variety of phenomena \cite{Chakraborty_2016,Capozzi_2022,Liu_2025PhRvD.111b3051L,Tamborra_2021,Richers_2022,Fischer_2024,Volpe_2024,Johns_2025}.

The so-called fast flavor conversion (FFC) is one of such collective oscillation modes that are induced by differences in the angular distributions of \(\nu_e\) and \(\bar{\nu}_e\) in momentum space. It occurs very rapidly (on the timescale shorter by many orders than the dynamical timescale of PNS) and many studies have demonstrated that FFC will occur rather commonly in the CCSN core \cite{Delfan_Azari_2019,Nagakura_2019,Nagakura_2021PhRvD.104h3025N,Capozzi_2021,Zaizen_2021,Abbar_2021,Cornelius_2025}. The possibility of FFC in the PNS in its cooling phase was also studied \cite{Akaho_2024,Zaizen_2024} but for non-rotating PNS's.

In this paper we will study FFC in our rotating models. Since our simulations do not incorporate the neutrino oscillations, we will take a post-process approach \cite{Akaho_2024}, which is actually a common practice (see also \cite{Nagakura_2024PhRvD.109h3013N,Xiong_2025PhRvL.134e1003X,Akaho_2025arXiv250607017A} for the implementation of FFC onto classical neutrino transport). We look into the energy-integrated angular distribution of the net electron-type neutrinos, \(\Delta G\), which is defined as follows and calculated from our simulation results:
\begin{equation} \label{eq:DeltaG}
\Delta G = \sqrt{2}G_F\int \frac{\epsilon^2d\epsilon}{2\pi^2}(f_{\nu_e}-f_{\bar{\nu}_e}).
\end{equation}

It is known that the presence of zero-crossings in \(\Delta G\) at some angles is the condition for FFC \cite{Nagakura_2019,Morinaga_2020,Harada_2022,Morinaga_2022PhRvD.105j1301M,Akaho_2024}. In the region, where this condition is satisfied, we calculate the local growth rate of FFC approximately according to the following formula \cite{Akaho_2024}:

\begin{equation}\label{eq:sigmaFFI}
\sigma_{FFI}=\sqrt{-\left(\int_{\Delta G>0}\frac{d\Omega}{4\pi}\Delta G\right)\left(\int_{\Delta G<0}\frac{d\Omega}{4\pi}\Delta G\right)}.
\end{equation}

\noindent The angular integral in the first (second) parenthesis is limited to the angular region, where \(\Delta G\) is positive (negative). The subscript FFI (the abbreviation of fast flavor instability) is used exchangeably with FFC in this paper although the former is mostly used in the context of linear stability analysis in the literature. Note that if FFC occurs at some point of time indeed, the subsequent flavor contents will be modified in principle. In the current paper, however, we ignore such feedbacks if any.

\begin{figure*}[htbp]
     \centering
     \begin{subfigure}[b]{0.325\textwidth}
         \centering
         \includegraphics[width=\linewidth]{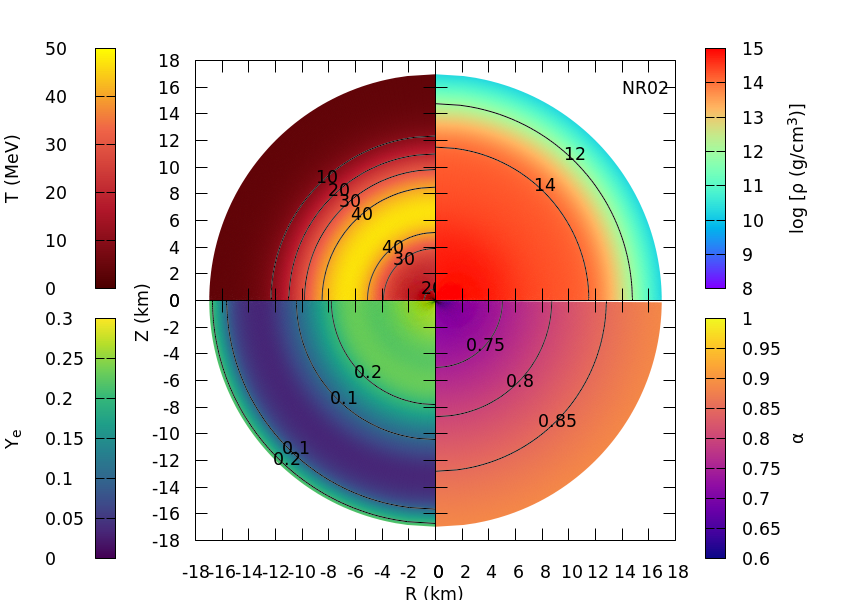}
     \end{subfigure}
     \begin{subfigure}[b]{0.325\textwidth}
     \centering
     \includegraphics[width=\linewidth]{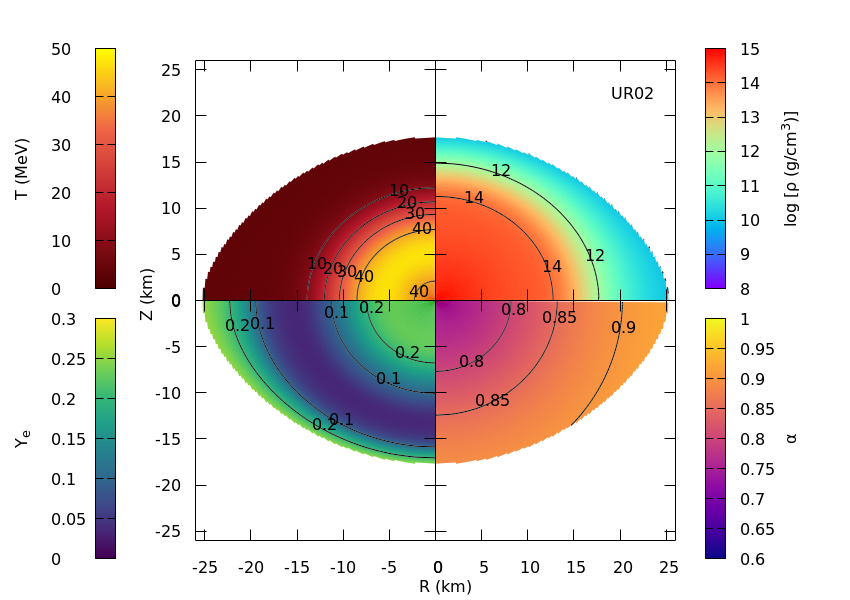}
     \end{subfigure}
     \begin{subfigure}[b]{0.325\textwidth}
         \centering
         \includegraphics[width=\linewidth]{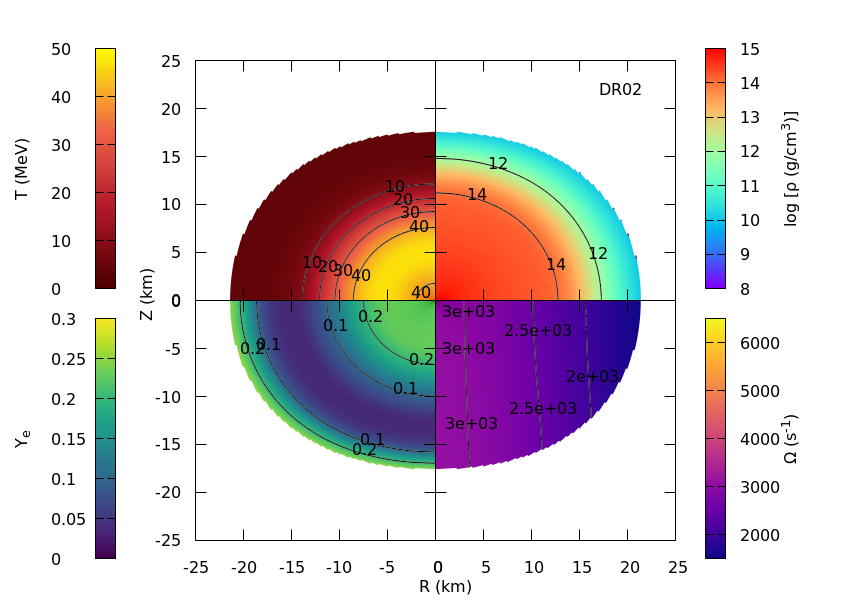}
     \end{subfigure}
     \centering
     \begin{subfigure}[b]{0.325\textwidth}
         \centering
         \includegraphics[width=\linewidth]{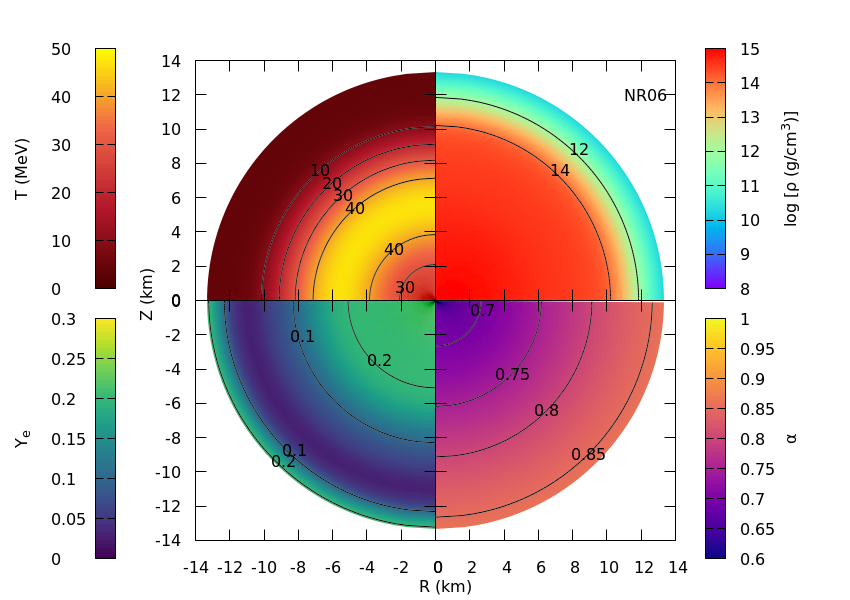}
     \end{subfigure}
     \begin{subfigure}[b]{0.325\textwidth}
     \centering
     \includegraphics[width=\linewidth]{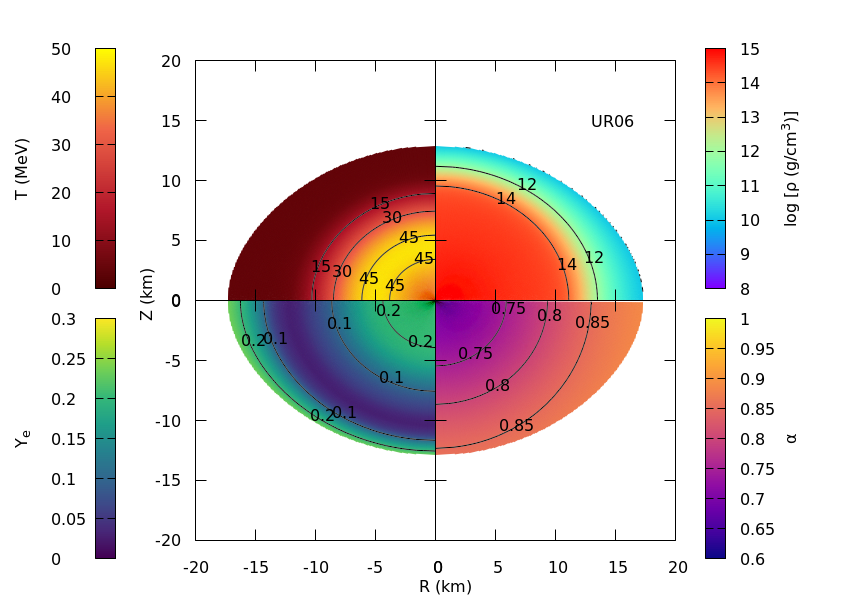}
     \end{subfigure}
     \begin{subfigure}[b]{0.325\textwidth}
         \centering
         \includegraphics[width=\linewidth]{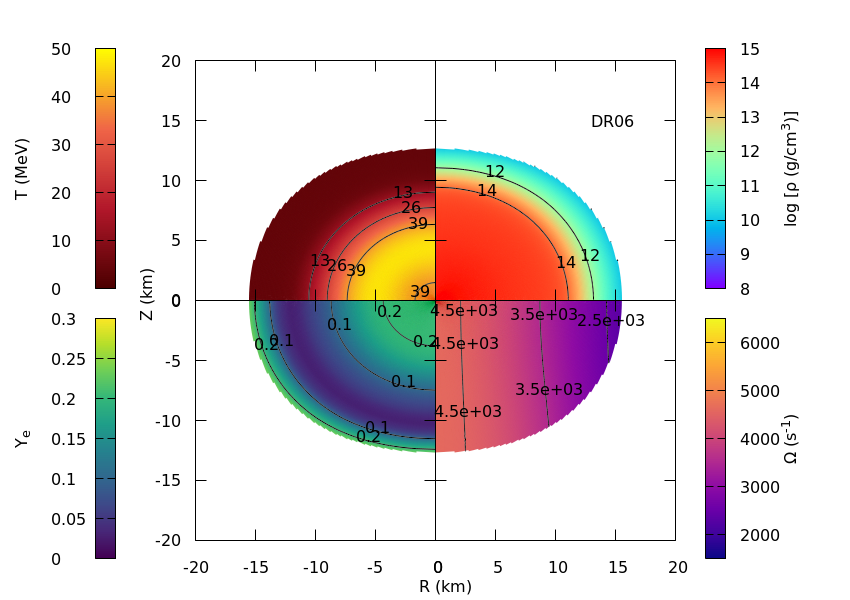}
     \end{subfigure}
     \centering
     \begin{subfigure}[b]{0.325\textwidth}
         \centering
         \includegraphics[width=\linewidth]{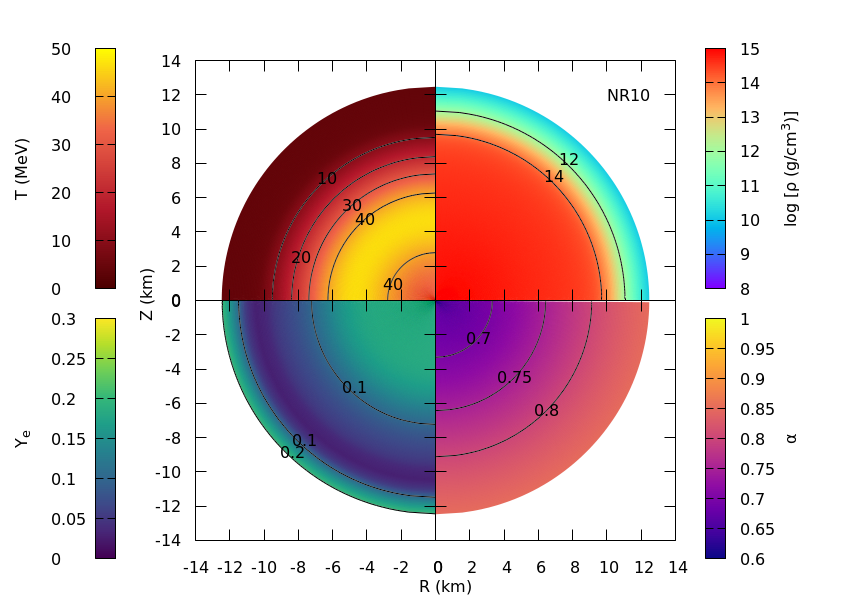}
     \end{subfigure}
     \begin{subfigure}[b]{0.325\textwidth}
     \centering
     \includegraphics[width=\linewidth]{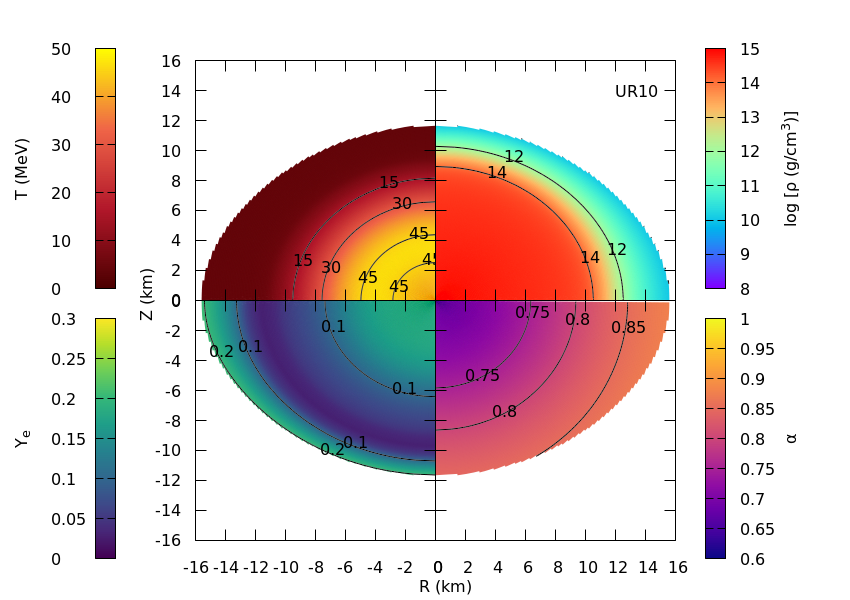}
     \end{subfigure}
     \begin{subfigure}[b]{0.325\textwidth}
         \centering
         \includegraphics[width=\linewidth]{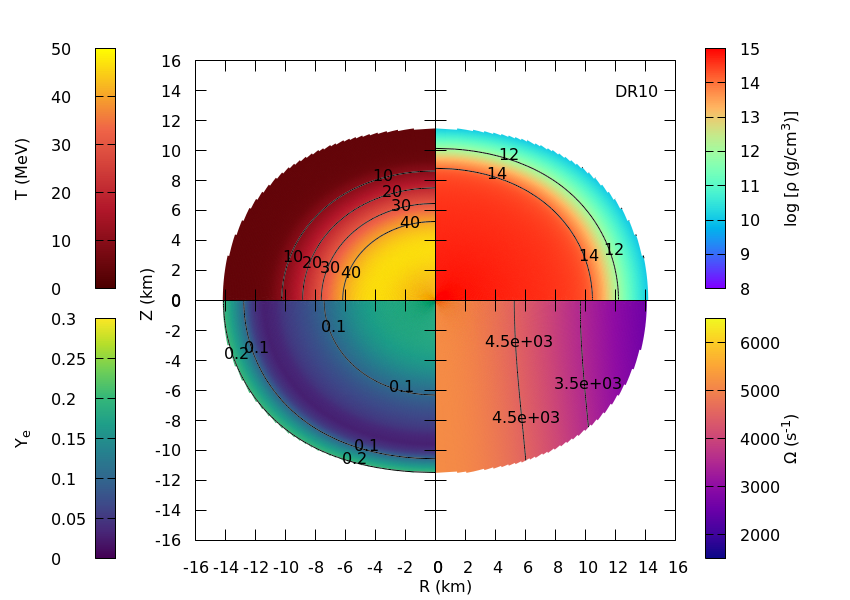}
     \end{subfigure}
     \centering
     \begin{subfigure}[b]{0.325\textwidth}
         \centering
         \includegraphics[width=\linewidth]{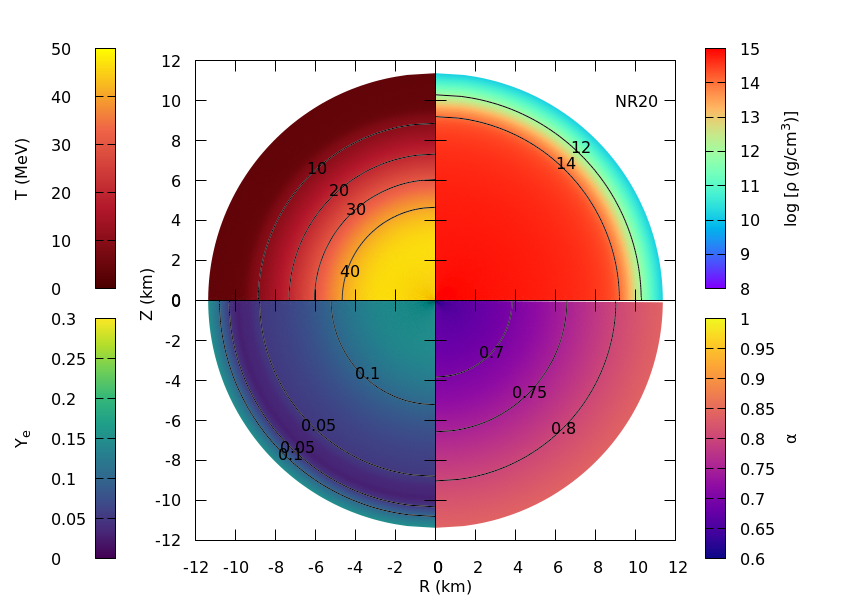}
     \end{subfigure}
     \begin{subfigure}[b]{0.325\textwidth}
     \centering
     \includegraphics[width=\linewidth]{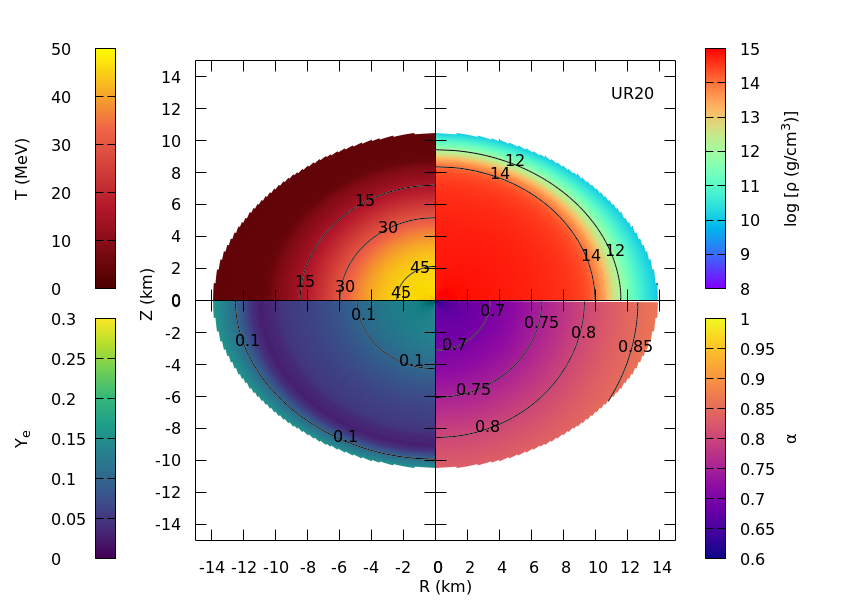}
     \end{subfigure}
     \begin{subfigure}[b]{0.325\textwidth}
         \centering
         \includegraphics[width=\linewidth]{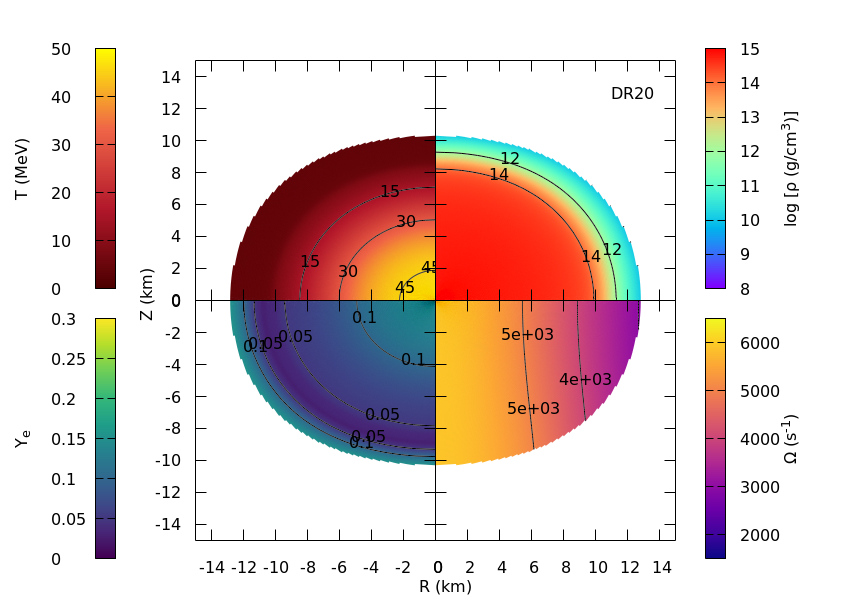}
     \end{subfigure}
     \centering
     \begin{subfigure}[b]{0.325\textwidth}
         \centering
         \includegraphics[width=\linewidth]{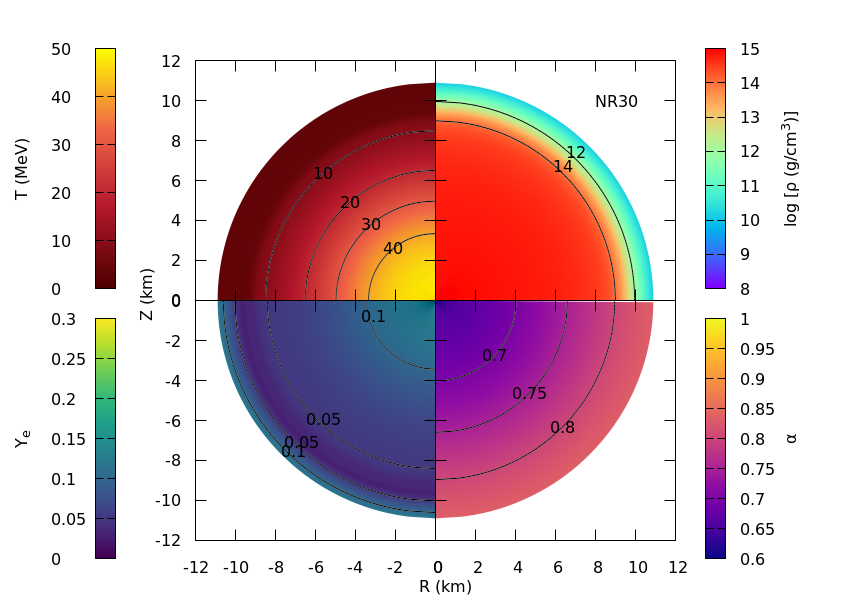}
     \end{subfigure}
     \begin{subfigure}[b]{0.325\textwidth}
     \centering
     \includegraphics[width=\linewidth]{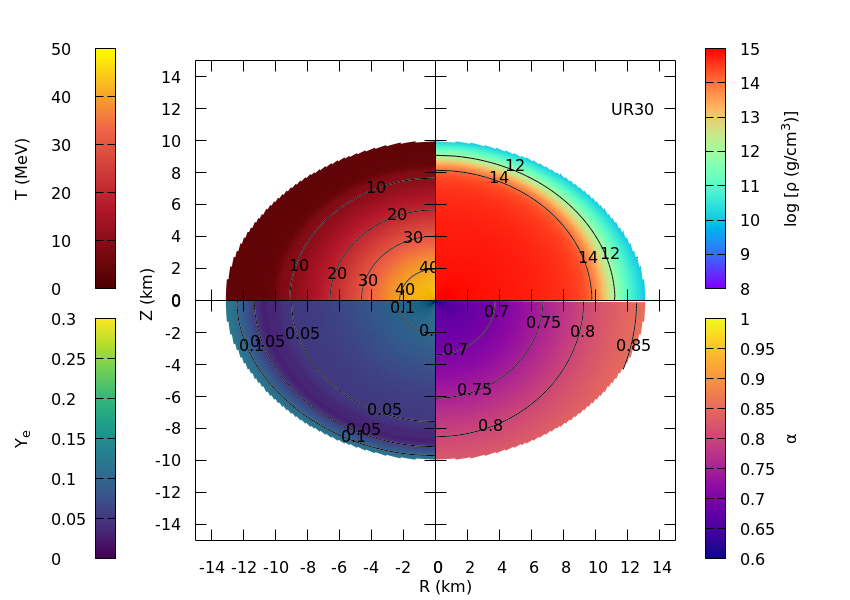}
     \end{subfigure}
     \begin{subfigure}[b]{0.325\textwidth}
         \centering
         \includegraphics[width=\linewidth]{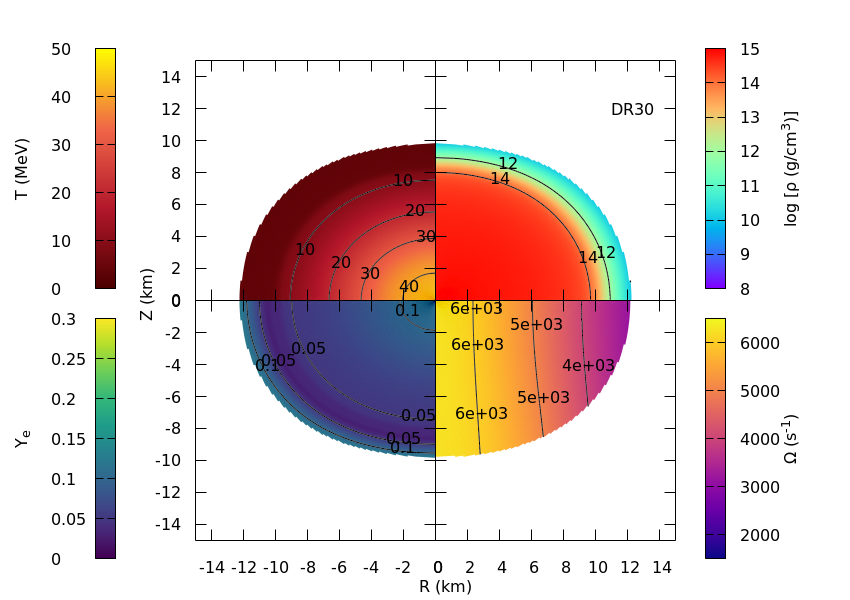}
     \end{subfigure}
     \caption{2D profiles with contours of the temperature (top left quadrant), density (top right quadrant), electron fraction (bottom left quadrant) and lapse function for non-rotating and uniformly-rotating models or angular velocity for the differentially-rotating model (bottom right quadrant) at \(t=2\)s (first row), \(6\)s (second row), \(10\)s (third row), \(20\)s (fourth row) and \(30\)s (fifth row) for the non-rotating (left column), rigidly-rotating (central column) and differentially-rotating (right column) models.}
     \label{fig:background}
\end{figure*}

\section{Results}\label{Res}

Now the main results are presented. After summarizing the background models in Section \ref{Res-Basics} we we analyze in Section~\ref{Res-Nulum} the anisotropies in the neutrino luminosities and energy spectra quantitatively in detail, based on the shape of, and the temperature and \(Y_e\) distributions on the neutrino-hemisphere defined observer-wise (see Eq. \ref{eq:obs-dependentLum} in Section \ref{NumMeth-Nulum}). In Section~\ref{Res-GW} we show the results regarding the gravitational waves (GW) emitted in association with the anisotropic neutrino radiations. Finally in Section \ref{Res-FFI} we discuss the possibility of FFC by looking for the zero-crossing in the energy-integrated angular distribution of the net electron-type neutrinos (Eq. \ref{eq:DeltaG} in Section \ref{NumMeth-FFI}).

\subsection{Background PNS models}\label{Res-Basics}
We first summarize the properties of our rotational PNS models. The information on some relevant quantities is given in Table \ref{tab:carquan} and the spatial distributions in (the quadrant of) the meridian section of the temperature,  density, electron fraction in addition to the lapse function of the spacetime and the angular velocity are shown as contour plots on top of the corresponding color maps in Fig. \ref{fig:background}. We consider two time-sequences, one in the uniform-rotation and the other in the differential-rotation (see Section \ref{NumMeth-rot}), in addition to the non-rotational sequence for reference. Along each sequence the baryonic mass and the total angular momentum are kept (nearly) constant so that they could be regarded as crude approximations to the actual evolutions. 

The two rotational sequences are in a sense two extremes: in the former the angular momentum is supposed to be transferred very efficiently among fluid elements to keep uniform-rotation whereas in the latter the specific angular momentum is almost constant at large cylindrical radii, meaning that the configuration is marginally Rayleigh-stable there. Note that the actual spatial distribution of angular momentum is determined by the PNS contraction via neutrino cooling and the angular momentum transfer by not-well-understood  mechanisms \cite{Margalit_2022}. It will change in time even if the angular momentum transfer is absent and the specific angular momentum is frozen to each fluid element. In this paper, as mentioned earlier, the matter profiles are fixed during the simulations, which typically last for a few milliseconds.

As found from Table \ref{tab:carquan}, all the models have rotation periods of \(\lesssim 5\)ms on the equatorial surface at the earliest time of \(t=2\)s. As the time passes, they cool and shrink via neutrino emissions and, as a result, they spin up. The rotation period becomes \(\lesssim 2\)ms and \(|T/W|\) rises up monotonically and exceeds \(0.05\) at \(t=30\)s in both sequences. The central density also increases monotonically.

As can be seen from Fig. \ref{fig:background}, the rapidly-rotating PNS's are substantially flattened due to centrifugal forces. The ratio of the polar radius to the equatorial radius, \(r_p/r_e\), is larger for the uniform-rotation models than for the differential-rotation models at equal times. This is because the outer region rotates more rapidly for the former than for the latter. It is interesting that the ratio is not much changed in time for both sequences despite the radii get smaller indeed. Note that the differential-rotation is rather limited to the equatorial surface region in our models (see the lower right quadrant of Fig. \ref{fig:background}). This reflects the expectation that the central region of the PNS, which is a descendant of the homologously-contracted inner core, will rotate nearly uniformly. As a result, the configurations of the inner part of the PNS's on the two sequences are not much different from each other at equal times. It is also reflected in the fact that the lapse functions are also very close to each other for the two sequences (and that is why it is shown only for the uniform-rotation models in Fig. \ref{fig:background}).

It should be emphasized, however, that the neutrino luminosities and energy spectra are formed near the surface of PNS and are hence affected by the rotations. In fact, compared with the non-rotational models, the rotational models have thicker low-temperature layers near the surface. The temperature is lower on the pole than on the equator at the same radius, or equivalently, the equi-temperature surfaces are oblate for both rotational models. The contrast is higher for the uniform-rotation models near the surface. The matter density is in general lower for the differential-rotation models although the difference is small near the center. As for the lapse function, the uniform-rotation models show slightly smaller values throughout the star.

\subsection{Observer-Dependent Neutrino Luminosity and Neutrino-Hemisphere}\label{Res-Nulum}

\begin{figure*}
     \centering
     \begin{subfigure}[b]{0.31\textwidth}
         \centering
         \includegraphics[width=\linewidth]{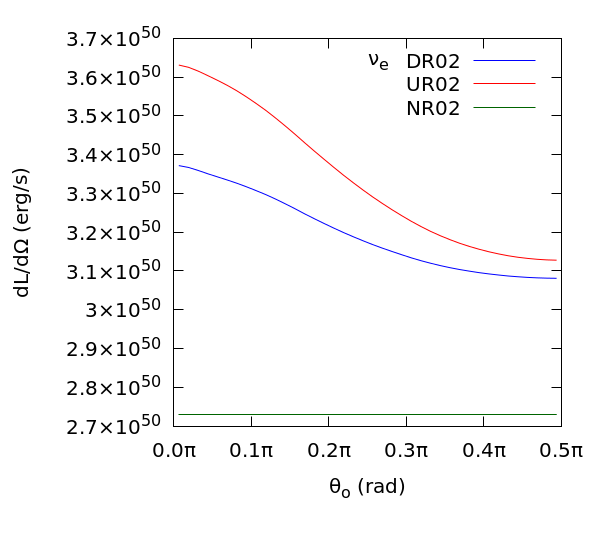}
     \end{subfigure}
     \begin{subfigure}[b]{0.31\textwidth}
         \centering
         \includegraphics[width=\linewidth]{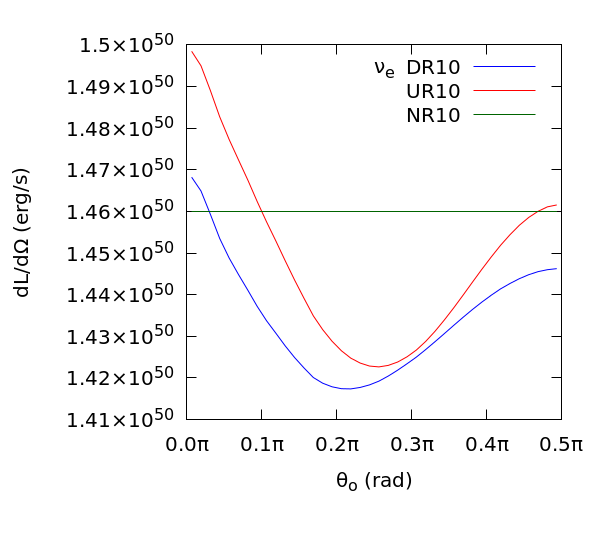}
     \end{subfigure}
     \begin{subfigure}[b]{0.31\textwidth}
     \centering
     \includegraphics[width=\linewidth]{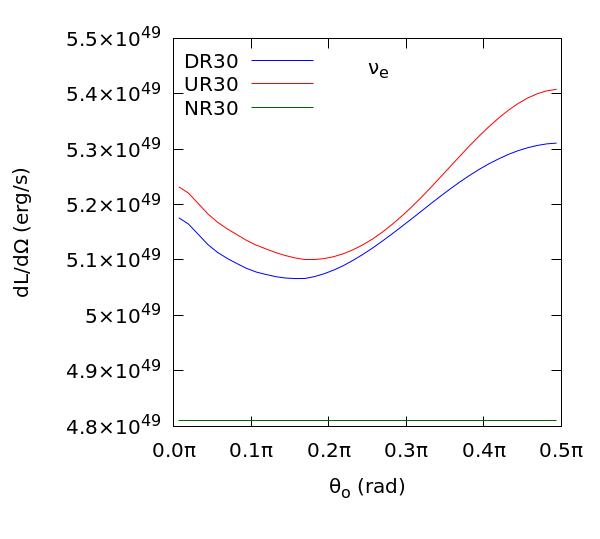}
     \end{subfigure}
     \centering
     \begin{subfigure}[b]{0.31\textwidth}
         \centering
         \includegraphics[width=\linewidth]{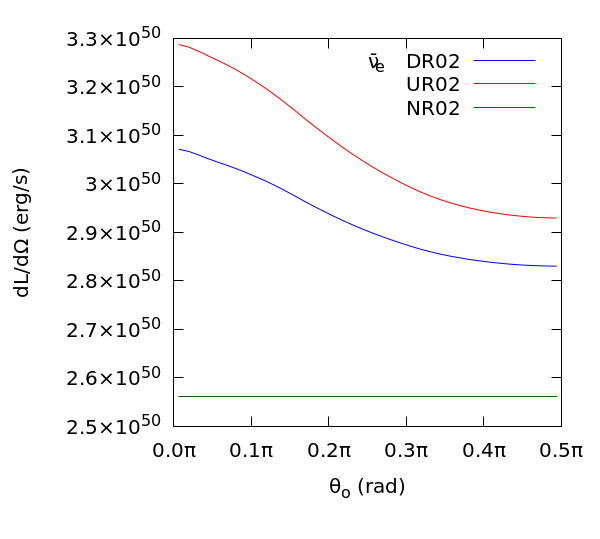}
     \end{subfigure}
     \begin{subfigure}[b]{0.31\textwidth}
         \centering
         \includegraphics[width=\linewidth]{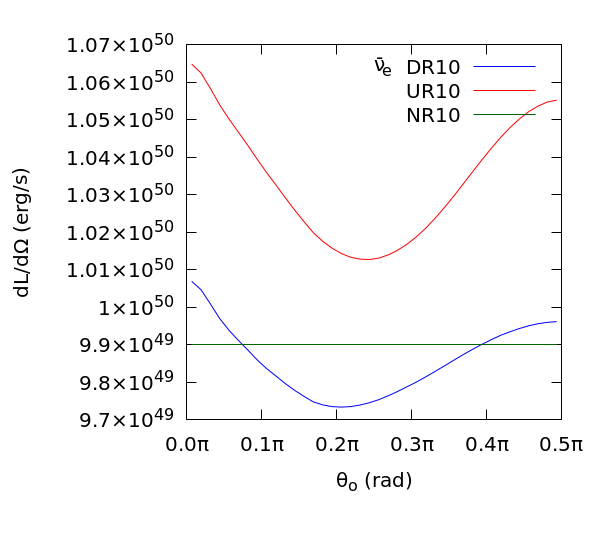}
     \end{subfigure}
     \begin{subfigure}[b]{0.31\textwidth}
     \centering
     \includegraphics[width=\linewidth]{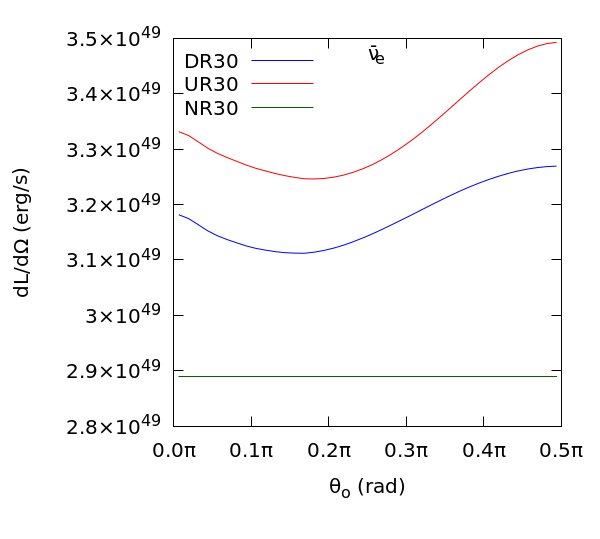}
     \end{subfigure}
     \centering
     \begin{subfigure}[b]{0.31\textwidth}
         \centering
         \includegraphics[width=\linewidth]{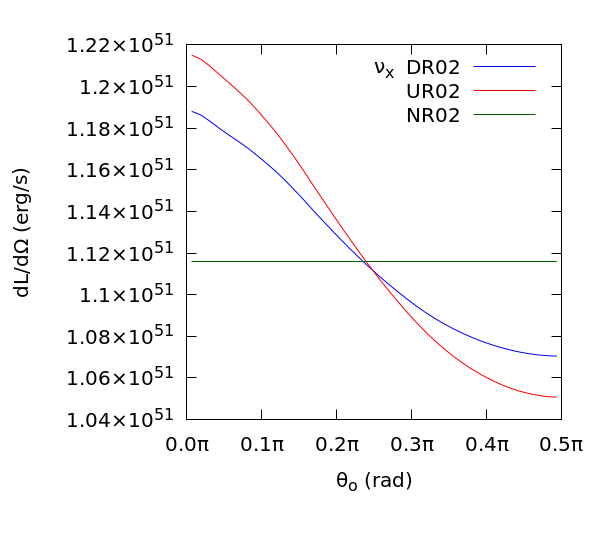}
     \end{subfigure}
     \begin{subfigure}[b]{0.31\textwidth}
         \centering
         \includegraphics[width=\linewidth]{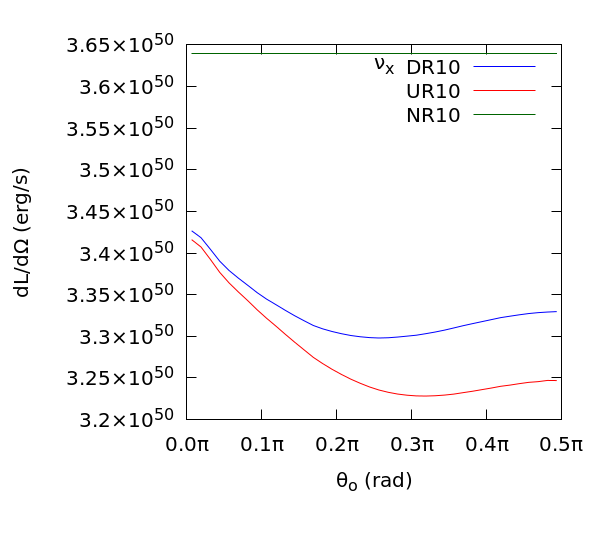}
     \end{subfigure}
     \begin{subfigure}[b]{0.31\textwidth}
     \centering
     \includegraphics[width=\linewidth]{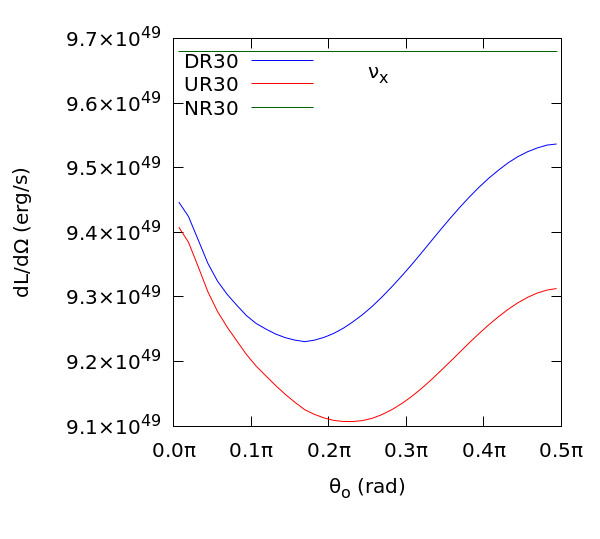}
     \end{subfigure}
     \caption{Observer-dependent luminosity as a function of the observer's position for \(\nu_e\) (top row), \(\bar{\nu}_e\) (central row) and \(\nu_x\) (bottom row) at \(t=2\)s (left column), \(10\)s (central column) and \(30\)s (right column) for the two rotational models. The non-rotational model is also plotted for reference.}
     \label{fig:dLdO}
\end{figure*}

Now we proceed to the main results, i.e., the anisotropy in the neutrino emissions. We look first at the observer-dependent neutrino luminosities defined as Eq. \ref{eq:obs-dependentLum}. We show them for each neutrino species separately in Fig. \ref{fig:dLdO} as functions of the zenith angle of the observer position at \(t=2\), \(10\), \(30\)s for the two rotational models. All three species show similar behavior: at the earliest time of \(t=2\)s, the luminosity is highest for the observer located on the pole and decreases monotonically with the zenith angle; as the time passes, a dip starts to develop at an intermediate angle; it moves towards the pole with time; at the same time the luminosity on the equator becomes dominant over that on the pole. The anisotropy, i.e., the deviation from the average of \(dL/d\Omega\), is greatest at the earliest time, \(\sim 16\%\) for the uniform-rotation model and \(\sim 9\%\) for the differential-rotation mode; it decreases down to \(\sim 5\% \) and \(\sim 3\%\), respectively, at \(t = 10\)s and rises slightly to \(\sim 6\%\) and \(\sim 5\%\), respectively, at \(t=30\)s. These features are further analyzed below.

\begin{figure}
    \centering
    \includegraphics[width=1\linewidth]{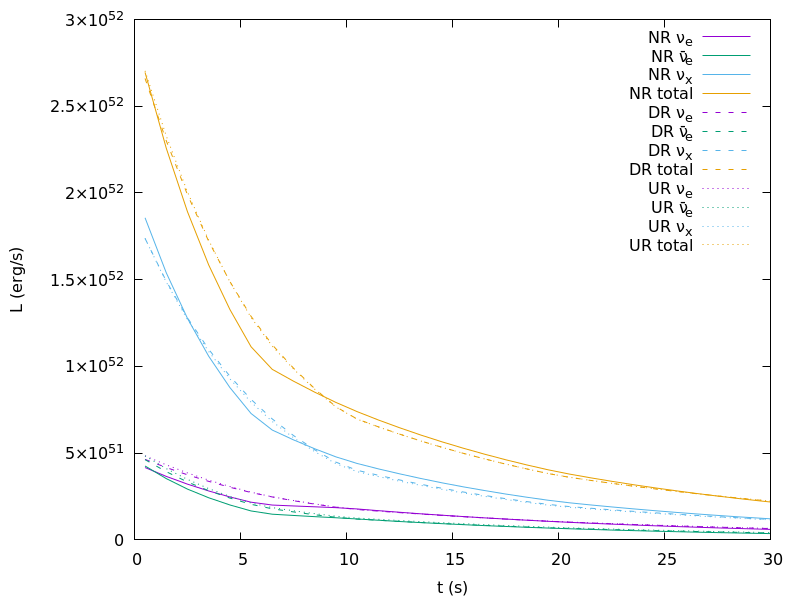}
    \caption{Total luminosity as a function of time for the three different models and the different neutrino species.}
    \label{fig:L_all}
\end{figure}

Before doing so, we compare the angle-integrated luminosities, i.e., the ordinary luminosities as functions of time for the different models and different neutrino species in Fig. \ref{fig:L_all}. As the PNS evolves, the luminosity decreases monotonically in this phase. Electron-type neutrinos are dominant for all models at all times. Comparing with the non-rotational models, we see a general tendency that the luminosities of \(\nu_e\) and \(\bar{\nu}_e\) are enhanced by rotation whereas that of \(\nu_{x}\) is lowered. Our non-rotating model is consistent with the model in \cite{Sugiura_2022,Nakazato_2013}, from which the matter profiles are extracted. The angle-integrated luminosities for the individual neutrino species as well as the sum thereof are slightly greater than the luminosities reported in other studies \cite{Fischer_2009,Fischer_2020}.

In order to understand more in detail the anisotropies found in the observer-dependent luminosities, we employ the observer-dependent neutrino-hemisphere defined in section \ref{NumMeth-Nulum}. It was defined as a set of points in PNS that have the optical depth equal to \(2/3\) along the line directed to the observer. In rotating PNS's, they depend on the observer position. We may think that neutrinos are emitted effectively from this hemisphere and hence that its shape and the average neutrino energy on it are two important factors in understanding neutrino luminosities and energy spectra at the observer's position. Since neutrino-matter interactions are energy- and species-dependent, so is the neutrino-hemisphere. In the following, we employ the average neutrino energies given as

\begin{equation}
    \langle E \rangle=\frac{\int f\epsilon^3 d\Omega_{\nu}d\epsilon}{\int f\epsilon^2 d\Omega_{\nu}d\epsilon},
\end{equation}

\noindent to determine the location of the neutrino-hemisphere. It is evaluated at \(r= 40\)km for each radial ray to the observer position in each model (see Fig \ref{fig:nu-sph-calc1}). Instead of the values thus obtained, however, we use the on-grid values closest to them for numerical simplicity. This may be justified, since we are interested in the understanding of the anisotropy of the observer-dependent neutrino luminosities. The values actually employed are given in Table \ref{tab:meanE} for readers' reference. They are similar to those found in previous studies \cite{Fischer_2010,Muller_2012}.

\begin{table}
    \centering
    \begin{tabular}{|c|c|c|c|}
         \hline
         Model & \(\langle E \rangle_{\nu_e}\) (MeV)  & \(\langle E \rangle_{\bar{\nu}_e}\) (MeV) & \(\langle E \rangle_{\nu_x}\) (MeV) \\
         \hline
         NR02 & 11.7 & 11.7 & 14.8  \\
         NR06 & 11.7 & 11.7 & 11.7  \\
         NR10 & 11.7 & 11.7 & 11.7  \\
         NR20 & 8.7 & 11.7 & 11.7  \\
         NR30 & 8.7 & 11.7 & 8.7  \\
         \hline
         UR02 & 11.7 & 11.7 & 11.7 \\
         UR06 & 11.7 & 11.7 & 11.7 \\
         UR10 & 11.7 & 11.7 & 11.7 \\
         UR20 & 11.7 & 11.7 & 11.7 \\
         UR30 & 11.7 & 11.7 & 8.7 \\
         \hline
         DR02 & 11.7 & 11.7 & 11.7 \\
         DR06 & 11.7 & 11.7 & 11.7 \\
         DR10 & 11.7 & 11.7 & 11.7 \\
         DR20 & 11.7 & 11.7 & 11.7 \\
         DR30 & 11.7 & 11.7 & 8.7 \\
         \hline
    \end{tabular}
    \caption{Mean neutrino energies employed in locating the neutrino-hemisphere for all models and neutrino species.}
    \label{tab:meanE}
\end{table}

\begin{figure*}
     \centering
     \begin{subfigure}[b]{0.32\textwidth}
         \centering
         \includegraphics[width=\linewidth]{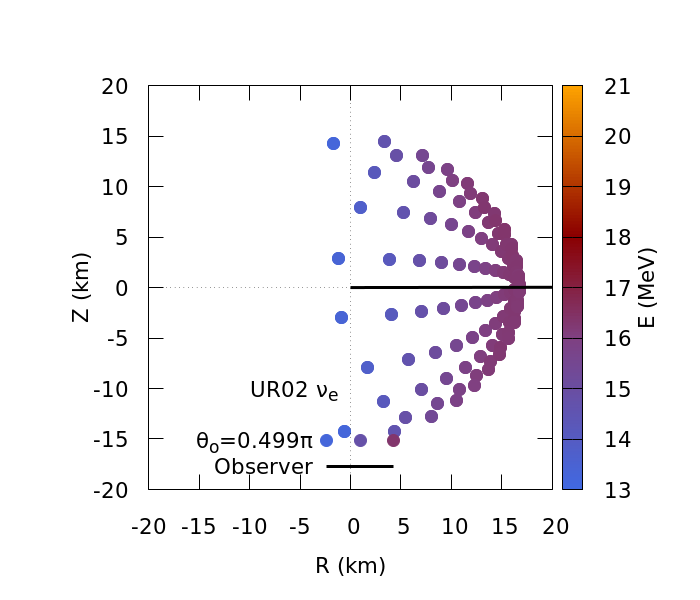}
     \end{subfigure}
     \begin{subfigure}[b]{0.32\textwidth}
         \centering
         \includegraphics[width=\linewidth]{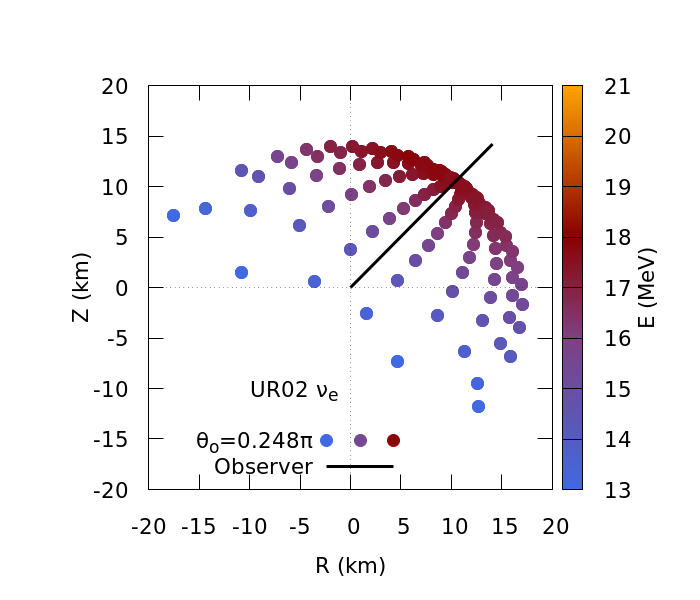}
     \end{subfigure}
     \begin{subfigure}[b]{0.32\textwidth}
         \centering
         \includegraphics[width=\linewidth]{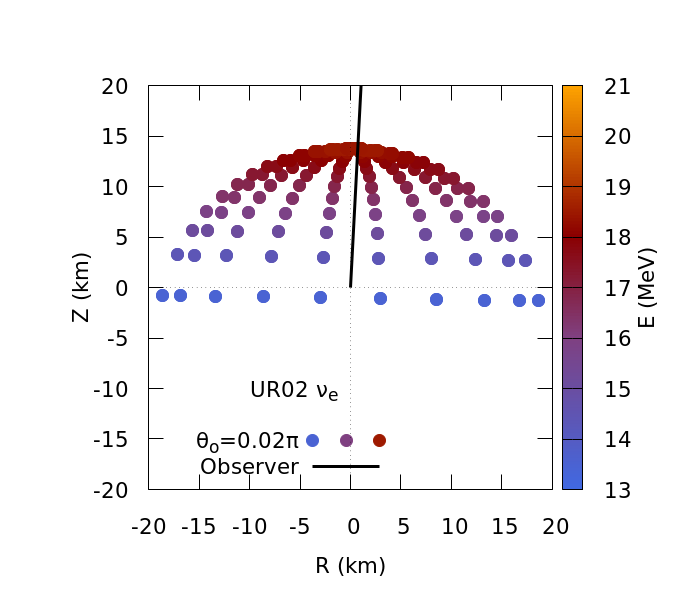}
     \end{subfigure}
     \centering
     \begin{subfigure}[b]{0.32\textwidth}
         \centering
         \includegraphics[width=\linewidth]{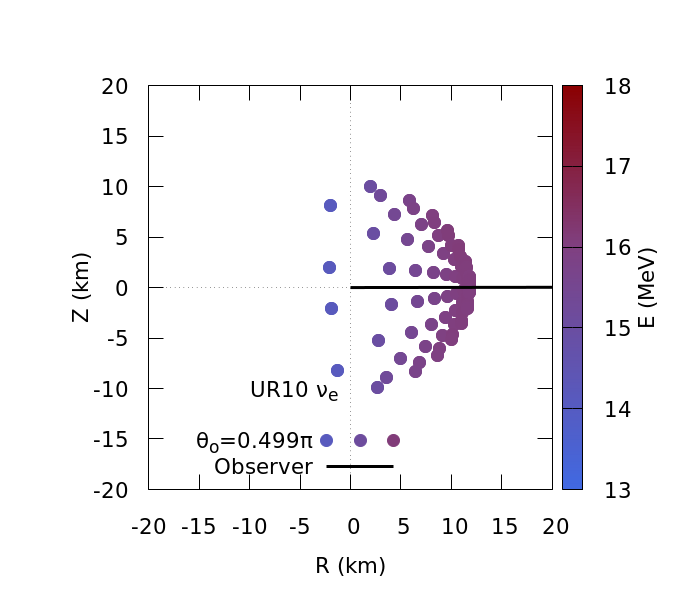}
     \end{subfigure}
     \begin{subfigure}[b]{0.32\textwidth}
         \centering
         \includegraphics[width=\linewidth]{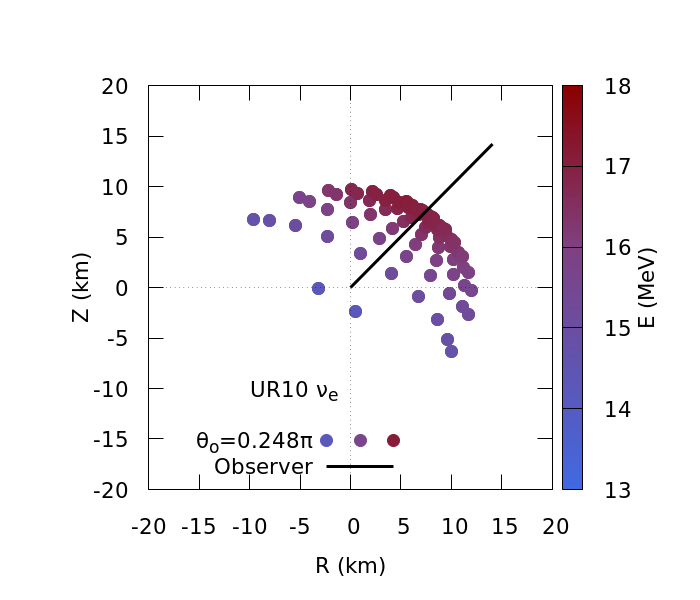}
     \end{subfigure}
     \begin{subfigure}[b]{0.32\textwidth}
         \centering
         \includegraphics[width=\linewidth]{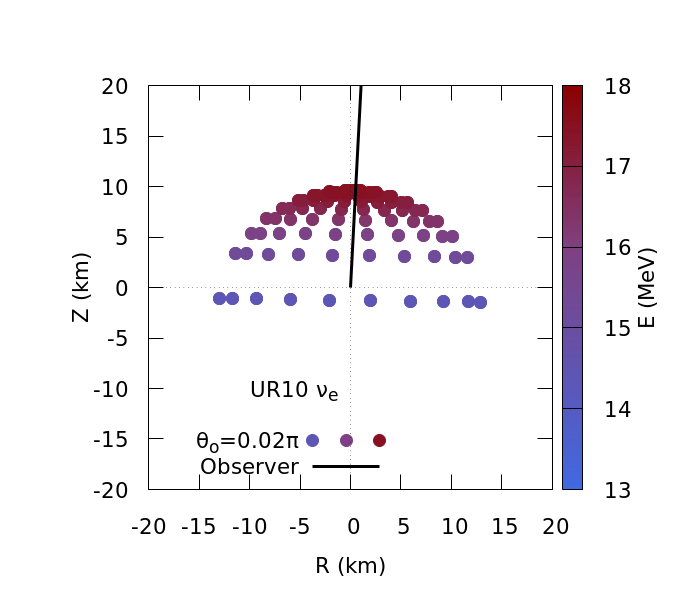}
     \end{subfigure}
     \centering
     \begin{subfigure}[b]{0.32\textwidth}
         \centering
         \includegraphics[width=\linewidth]{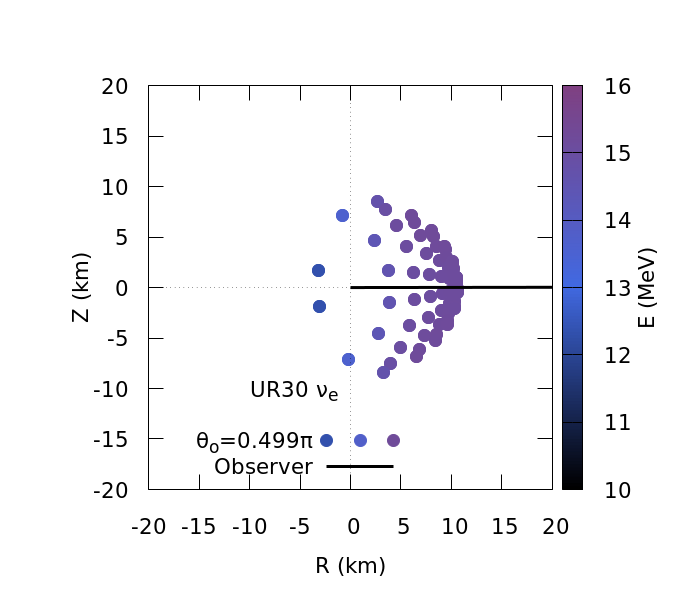}
     \end{subfigure}
     \begin{subfigure}[b]{0.32\textwidth}
         \centering
         \includegraphics[width=\linewidth]{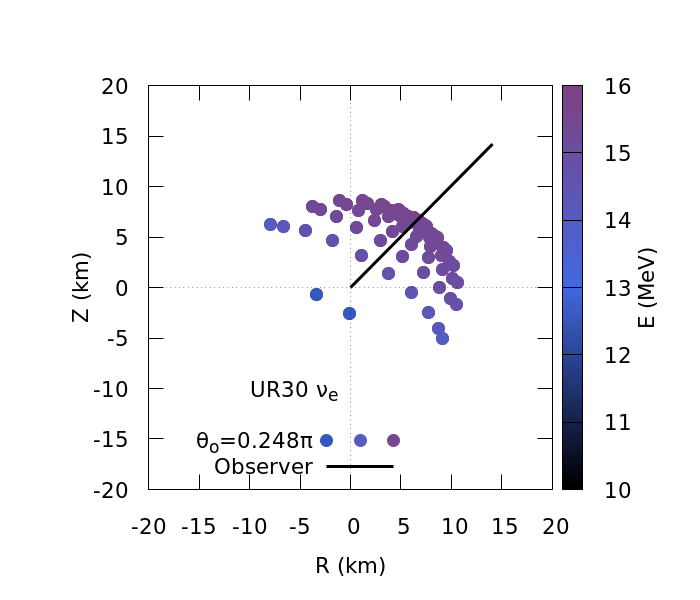}
     \end{subfigure}
     \begin{subfigure}[b]{0.32\textwidth}
         \centering
         \includegraphics[width=\linewidth]{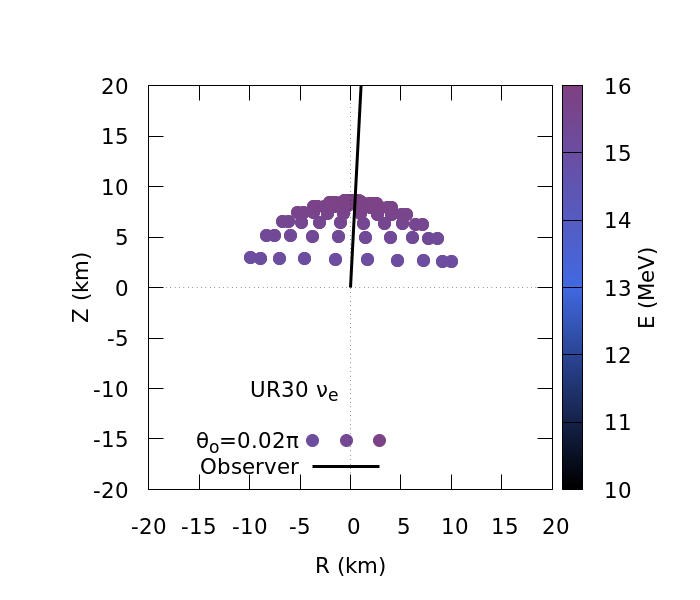}
     \end{subfigure}
        \caption{Neutrino-hemisphere position for the equator-observer (left panels), intermediate-observer (middle panels) and pole-observer (right panels). Observers with the mean neutrino energy of \(\nu_{e}\) in color. The top, middle and bottom rows correspond to models UR02, UR10 and UR30, respectively.}
        \label{fig:nusph-URnue-meanE}
\end{figure*}

\begin{figure*}
     \centering
     \begin{subfigure}[b]{0.32\textwidth}
         \centering
         \includegraphics[width=\linewidth]{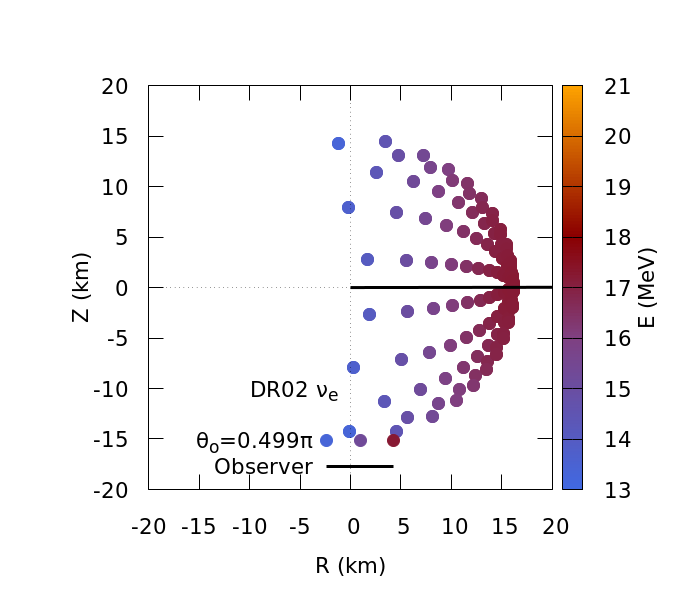}
     \end{subfigure}
     \begin{subfigure}[b]{0.32\textwidth}
         \centering
         \includegraphics[width=\linewidth]{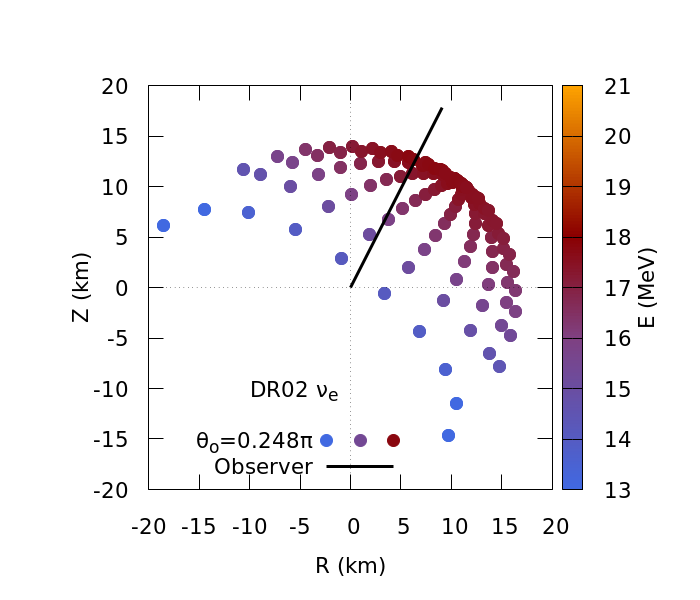}
     \end{subfigure}
     \begin{subfigure}[b]{0.32\textwidth}
         \centering
         \includegraphics[width=\linewidth]{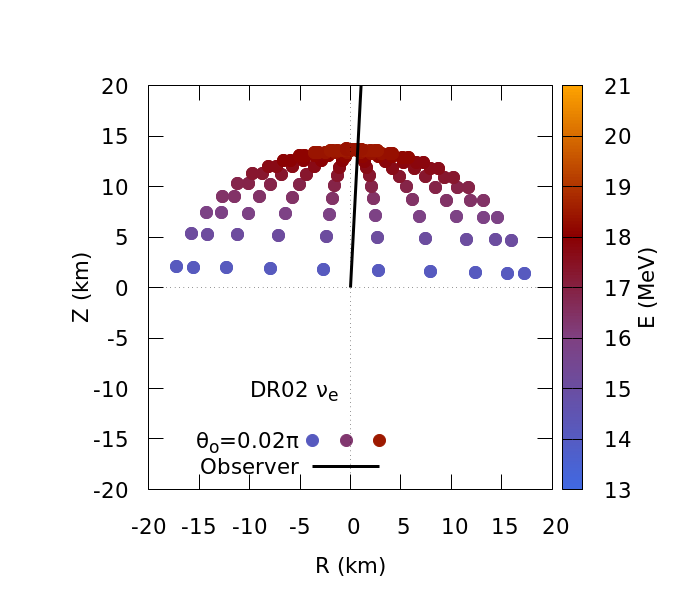}
     \end{subfigure}
     \centering
     \begin{subfigure}[b]{0.32\textwidth}
         \centering
         \includegraphics[width=\linewidth]{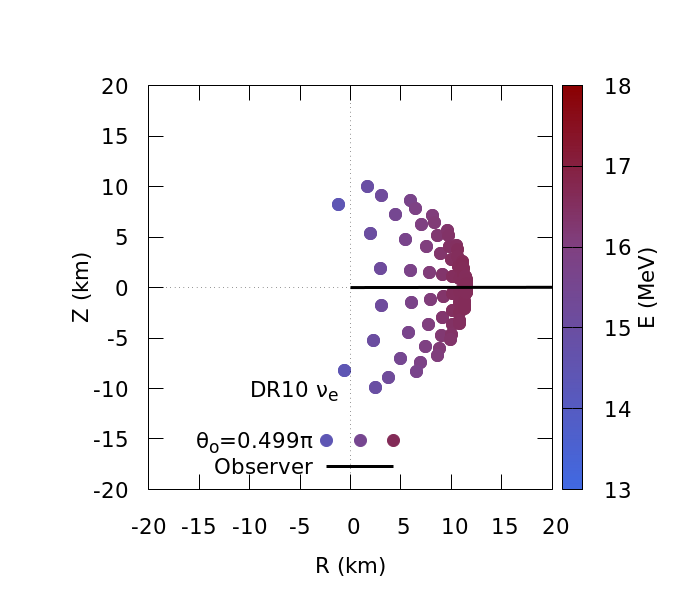}
     \end{subfigure}
     \begin{subfigure}[b]{0.32\textwidth}
         \centering
         \includegraphics[width=\linewidth]{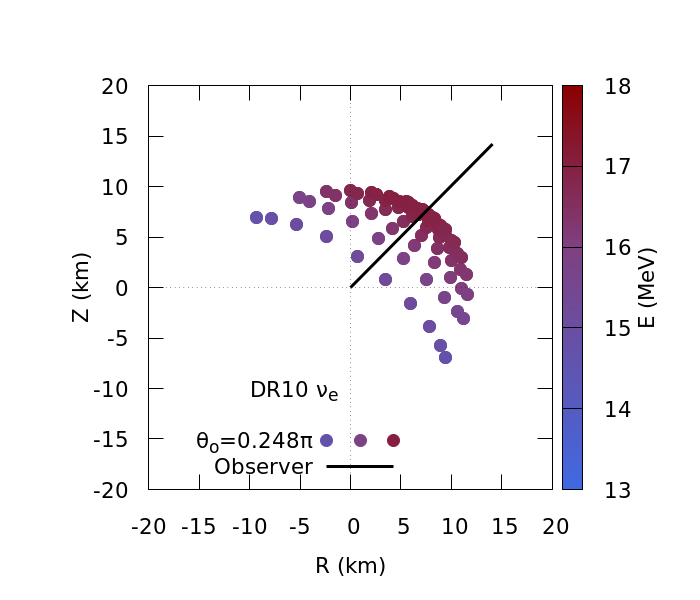}
     \end{subfigure}
     \begin{subfigure}[b]{0.32\textwidth}
         \centering
         \includegraphics[width=\linewidth]{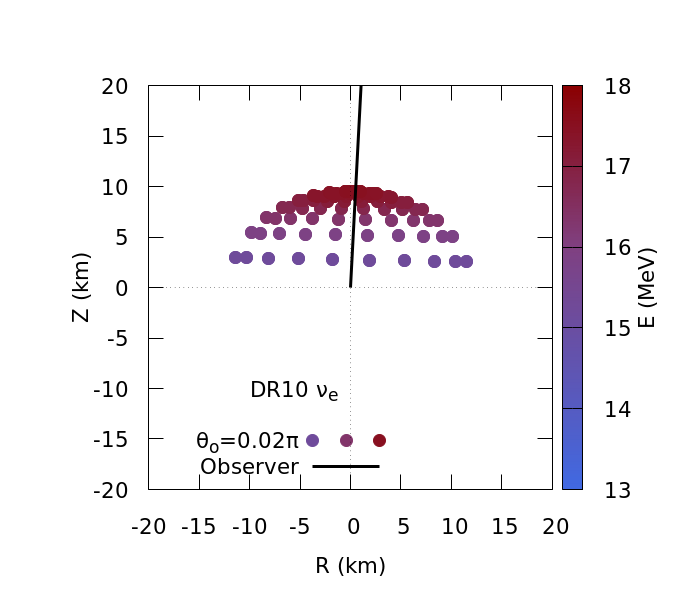}
     \end{subfigure}
     \centering
     \begin{subfigure}[b]{0.32\textwidth}
         \centering
         \includegraphics[width=\linewidth]{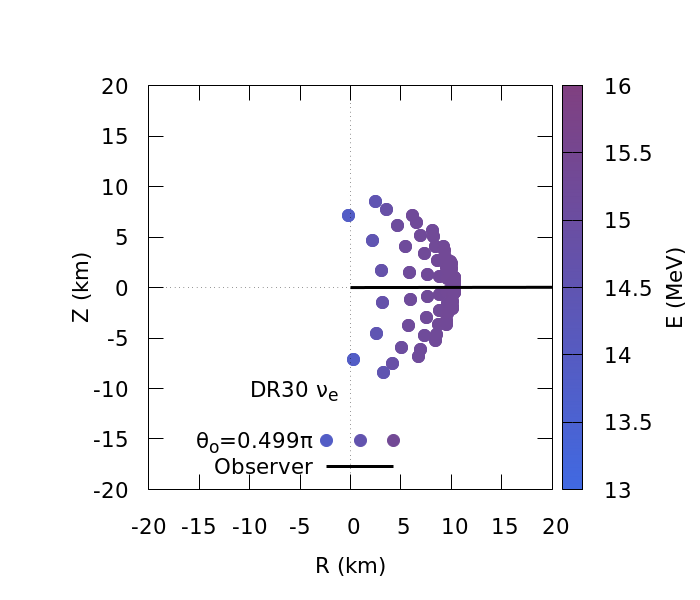}
     \end{subfigure}
     \begin{subfigure}[b]{0.32\textwidth}
         \centering
         \includegraphics[width=\linewidth]{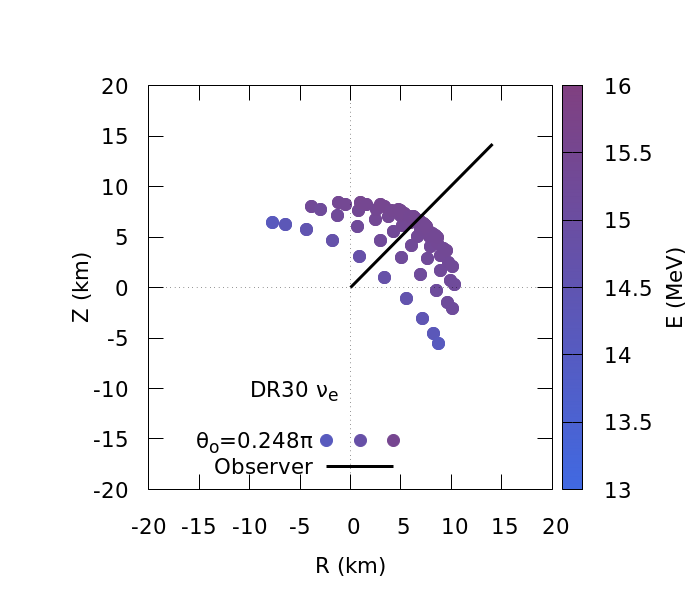}
     \end{subfigure}
     \begin{subfigure}[b]{0.32\textwidth}
         \centering
         \includegraphics[width=\linewidth]{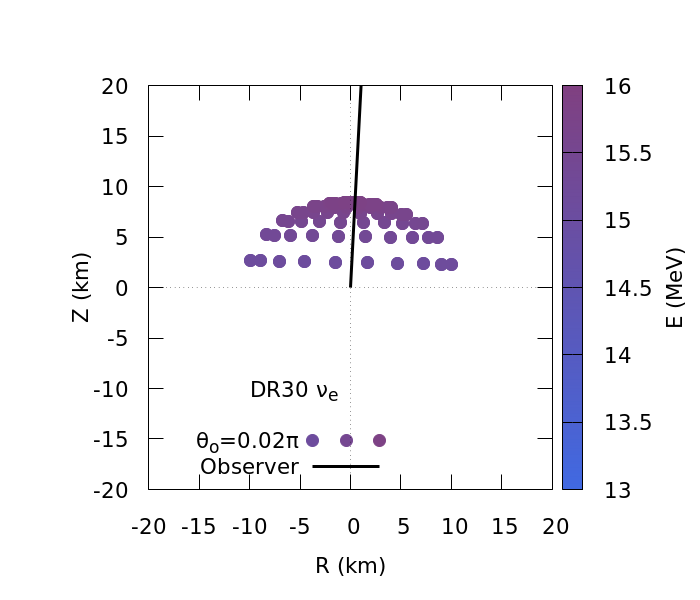}
     \end{subfigure}
        \caption{Same as Fig. \ref{fig:nusph-URnue-meanE} but for DR models.}
        \label{fig:nusph-DRnue-meanE}
\end{figure*}

We show in Figs. \ref{fig:nusph-URnue-meanE} and \ref{fig:nusph-DRnue-meanE} the position of the neutrino-hemisphere and the mean neutrino energies on it at \(t=2\)s, \(10\)s and \(30\)s for the uniform- and differential-rotation models, respectively. They are the sideviews seen from the negative \(y\)-axis (or the projection onto the \(x-z\) plane). The solid black line shows the observer direction in each case. Each dot corresponds to the sampling rays in Fig. \ref{fig:nu-sph-calc1}; the positions of these dots represent the position of the neutrino-hemisphere, where the optical depth is \(2/3\) from infinity (actually from the radius of \(r=40\)km) along those rays; the color of the dot shows the value of the mean neutrino energy at its position on the neutrino-hemisphere. 

As can be seen from these figures, the size of the neutrino-hemisphere shrinks with time just as the PNS itself does. It is also apparent that they are not perfect hemispheres with a constant radius even for the non-rotating models. This is simply because the peripheral rays go through a low-density region of PNS and the optical depth reaches \(2/3\) at larger distances from the starting points. For the rotational models, the neutrino-hemisphere is skewed. This is again understood from their oblate configurations, in which matter is thicker near the equator than around the pole. Also note that the neutrino-hemisphere is more deformed at earlier times. This is consistent with the matter profiles given in Fig. \ref{fig:background}: as the PNS gets more compact in time, its oblateness is reduced for both rotational models. The mean neutrino energy is highest near the radial ray and gets lower as the periphery of the neutrino-hemisphere is approached. This is something analogous to the limb darkening. Again for the rotational models, the energy peak is shifted slightly toward the pole although it is sometimes difficult to see it in the plots. It is also difficult to discern from the figures but we find that the inequality, \(R_{\nu_x}<R_{\bar{\nu}_e}<R_{\nu_e}\), holds in general as it does in the PNS cooling calculations in spherical symmetry. The values of the neutrino-hemisphere radius we obtained are smaller a bit than those of the neutrino-sphere radius found in \cite{Fischer_2010} (see Fig. 15a in their paper). This may be due to the fact that their progenitor has a smaller mass (\(10M_{\odot}\)). In \cite{Fischer_2010} it is also shown that \( \overline{E}_x > \overline{E}_{\bar{e}} > \overline{E}_e\), which agrees with our findings.

\begin{figure*}
     \centering
     \begin{subfigure}[b]{0.325\textwidth}
         \centering
         \includegraphics[width=\linewidth]{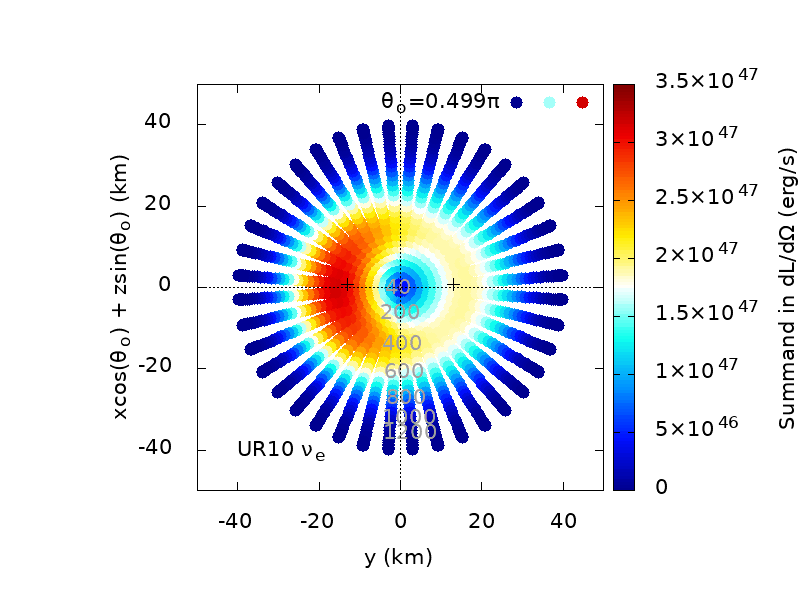}
     \end{subfigure}
     \begin{subfigure}[b]{0.325\textwidth}
         \centering
         \includegraphics[width=\linewidth]{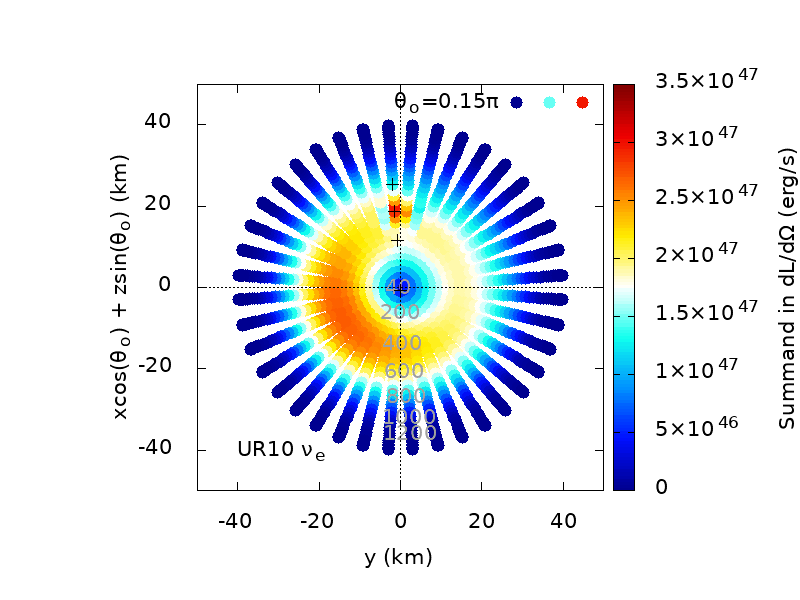}
     \end{subfigure}
     \begin{subfigure}[b]{0.325\textwidth}
         \centering
         \includegraphics[width=\linewidth]{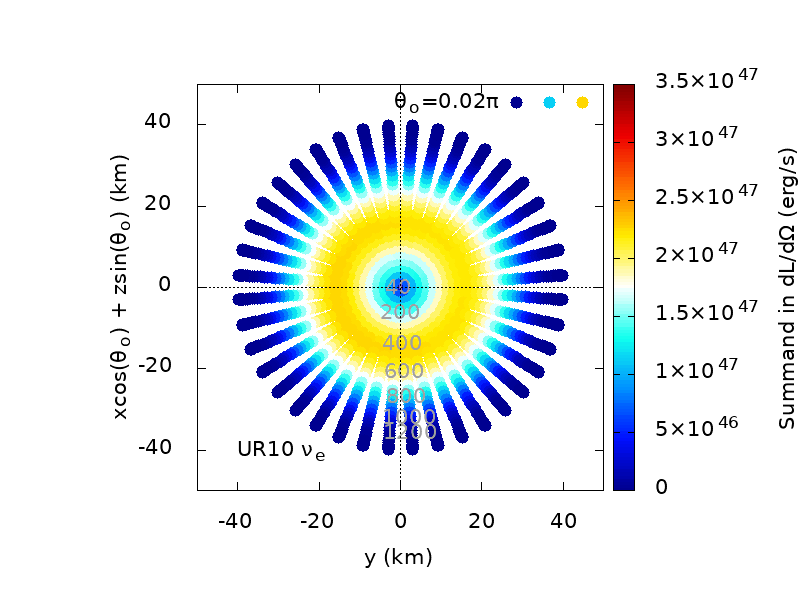}
     \end{subfigure}
     \centering
     \begin{subfigure}[b]{0.325\textwidth}
         \centering
         \includegraphics[width=\linewidth]{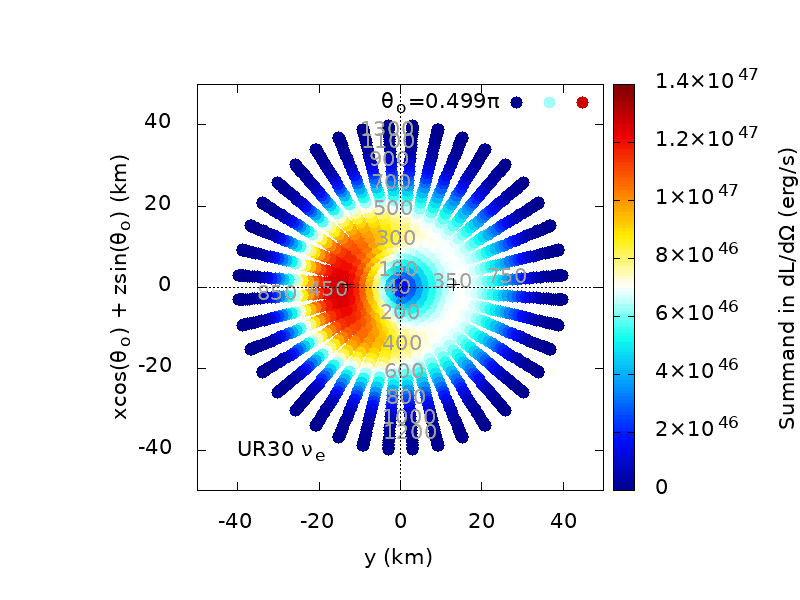}
     \end{subfigure}
     \begin{subfigure}[b]{0.325\textwidth}
         \centering
         \includegraphics[width=\linewidth]{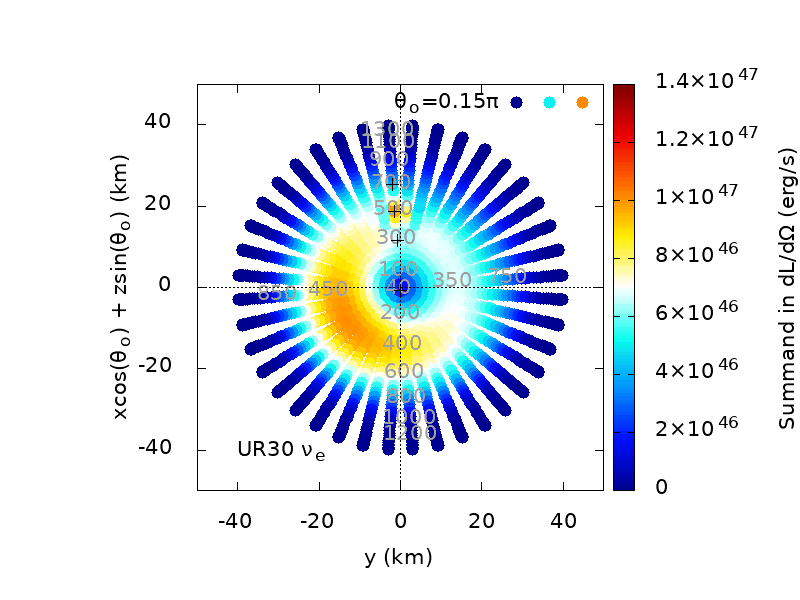}
     \end{subfigure}
     \begin{subfigure}[b]{0.325\textwidth}
         \centering
         \includegraphics[width=\linewidth]{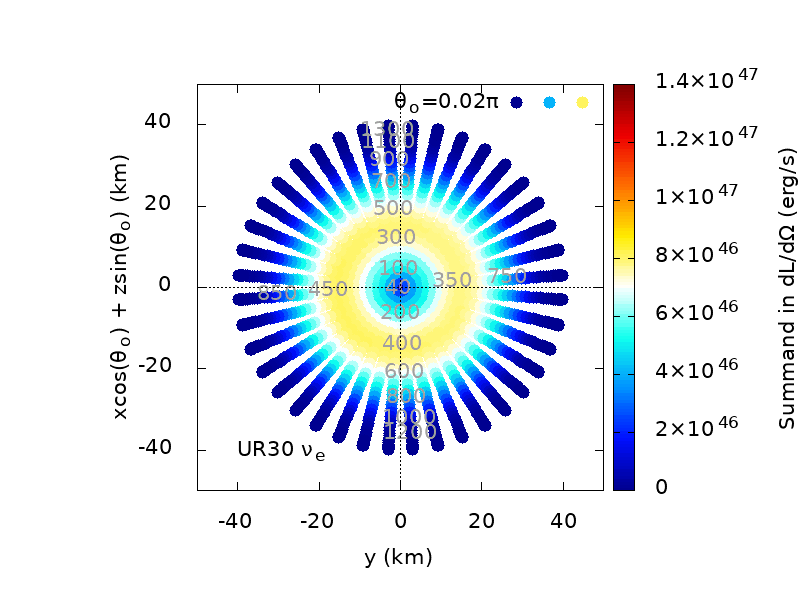}
     \end{subfigure}
        \caption{Summand of \(dL/d\Omega\) (Eq. \ref{eq:obs-dependentLum}) for \(\nu_e\) in models UR10 (top row) and UR30 (bottom row) for three different observer positions: the equator-observer (left panels), intermediate-observer (middle panels) and pole-observer (right panels). Each dot corresponds to one of the sampling points in Fig. \ref{fig:nu-sph-calc2}. They are numbered clockwise from the bottom point in the innermost circle outwards. So representative numbers are shown. All the summand values are evaluated at \(r=40\)km and displayed in color. The cross points in each of the left two panels indicates the points used for analysis of the red regions; the triple vertically-aligned cross points in each of the middle plots are used for analysis of the isolated red dot at the middle cross. See Fig. \ref{fig:cumulativesummands} and the text for details.}
        \label{fig:summands}
\end{figure*}

Although we find for the rotational models some offsets of the peak position in the mean neutrino energy from the radial ray, it seems unlikely to account for the appearance of the dip in \(dL/d\Omega\). We hence investigate \(dL/d\Omega\) itself in more detail. In Fig. \ref{fig:summands} we present the values of the integrand in Eq. \ref{eq:obs-dependentLum} for the electron-type neutrinos evaluated at individual sampling points on the sphere with the radius of \(40\)km; note that they are actually summands in the discretized equation and the integral measure is included. The upper and lower rows correspond to models UR10 and UR30, respectively; the three columns are for different observers as indicated with the values of \(\theta_o\) in each plot.

In the left column, the observer is assumed to be sitting almost on the equator (\(\theta_o = 0.499\pi\)). The north and south poles of PNS are seen from the side and correspond to the points on the vertical axis at \(\sim 20\)km from the origin. The equator of PNS is just the equator (or the horizontal axis) of the figure. The two points on the horizontal axis at \(\sim 25\)km from the origin are the equatorial points on PNS surface that are also seen from the side by the observer. In the right column, the observer is assumed to be located at \(\theta_o = 0.02\pi\), almost on the north pole. The origin of the figure corresponds to the north pole of PNS in this case. The circle with the radius of \(\sim 25\)km is the equator of PNS. Since our models are axisymmetric, the figure is circularly symmetric. The middle column corresponds to the observer located at \(\theta_o = 0.15\pi\), i.e., in between two cases. The north pole is sitting on the vertical axis but with a distance smaller than \(\sim 20\)km (in fact, corresponding to the red dot in the plot); the origin of the plot is the point at the latitude of \(0.15\pi\). The equator of PNS runs below the horizontal axis in this case. The north pole as well as (most of) the points on the PNS equator are seen from tilted directions. Only the point on the PNS surface that corresponds to the origin is viewed from the front by the observer. 

The color distribution in each plot are understood as follows. There are two  important things: (1) near the PNS surface the neutrino distributions are forward-peaked, i.e., they take the maximum value near \(\theta_{\nu}=0\), but (2) the peak direction is inclined from the local radial direction, \(\theta_{\nu}=0\). Since the neutrino-sphere is oblate in general, tracing the matter distribution elongated horizontally by centrifugal forces, the neutrino distribution is biased toward the rotation axis; it is also inclined in the azimuthal direction owing to the beaming effect of rotation. As a rule of thumb, the neutrino brightness is highest in the lower left region from the origin, as vindicated in the middle panels of Fig. \ref{fig:summands}.

\begin{figure}
     \centering
     \includegraphics[width=\linewidth]{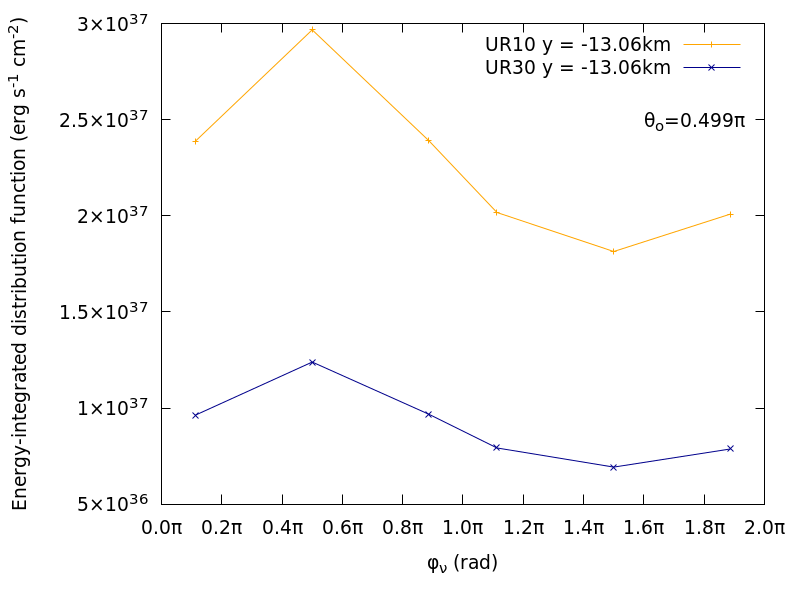}
     \caption{ Energy-integrated distribution function as a function of \(\phi_{\nu}\) at the cross point in the left panels of Fig. \ref{fig:summands}. The value of \(\theta_{\nu}\) is fixed to \(\theta_o\).}
    \label{fig:fvsphinu-leftright}
\end{figure}

Left-right asymetry is apparent in the left two panels: the left half is more reddish, implying that contributes more to \(dL/d\Omega\) than the right half. This asymmetry reflects the non-axisymmetry in the neutrino distributions in momentum space. In fact, their peaks are dislocated from the local radial direction because of the PNS rotation. Indeed, the distribution is inclined to the rotation direction. This is demonstrated in Fig. \ref{fig:fvsphinu-leftright}, in which we show the energy-integrated distribution functions at the points marked with a cross in the left panels of Fig. \ref{fig:summands}, which are roughly corresponding to the center of the reddish regions of the left plots in Fig. \ref{fig:summands}, as a function of \(\phi_{\nu}\); we fix \( \theta_{\nu}\) to \(\theta_o\). It is apparent that the distribution is indeed \( \phi_{\nu}\)-dependent and peaked at \(\phi_{\nu} = \pi/2\), which corresponds to the observer direction. Note that this is nothing but the Lorentz boost by rotation and also means that neutrinos carry angular momentum \cite{Harada_2019}. Since the peak directions are more aligned with the observer direction in the left half than the right half, the former is more reddish. Although these features are shared by the two plots for the different times, details are slightly different, reflecting the fact that rotation gets faster as the time passes and the PNS shrinks (Fig. \ref{fig:background}). Incidentally, the central regions are always bluish because the integral measure is smaller there.

For the intermediate-observer, whose location is close to the direction where the dip in \(dL/d\Omega\) occurs (Fig. \ref{fig:dLdO}), the picture is different as shown in the middle column: the lower left quadrant is more reddish. This is again due to the fact that the neutrino angular distributions in momentum space are non-axisymmetric and the peak directions are misaligned with the direction to the observer's direction as mentioned earlier. This time it is inclined not only to the rotation direction but also towards the rotation axis. This happens because the PNS is oblate by centrifugal forces; since neutrinos are emitted preferentially in the direction perpendicular to the PNS surface, which is deviated from the radial direction most at intermediate latitudes; for the oblate PNS, this implies the angular distribution mentioned above. This interpretation may be further supported by the fact that the dip becomes deeper at later times when the neutrino-hemisphere gets more skewed (see Fig. \ref{fig:nusph-URnue-meanE} and \ref{fig:nusph-DRnue-meanE}). As a result of the combination of this geometric effect with the rapid rotation of the PNS, the reddish region occurs in the lower left quadrant as observed. 

For the pole-observer, the picture is simplest as shown in the rightmost column for the two rotational models. Because of the (spatial) axisymmetry assumed in our models, the integrand of Eq. \ref{eq:obs-dependentLum} is circularly symmetric for this observer. It should be mentioned that although the neutrino distribution function itself has larger values near the pole than at other places, the color near the center is bluish because of the inclusion of the integral measure, which is smaller near the pole, in the integrand in the plots. 

\begin{figure}
    \centering
    \includegraphics[width=\linewidth]{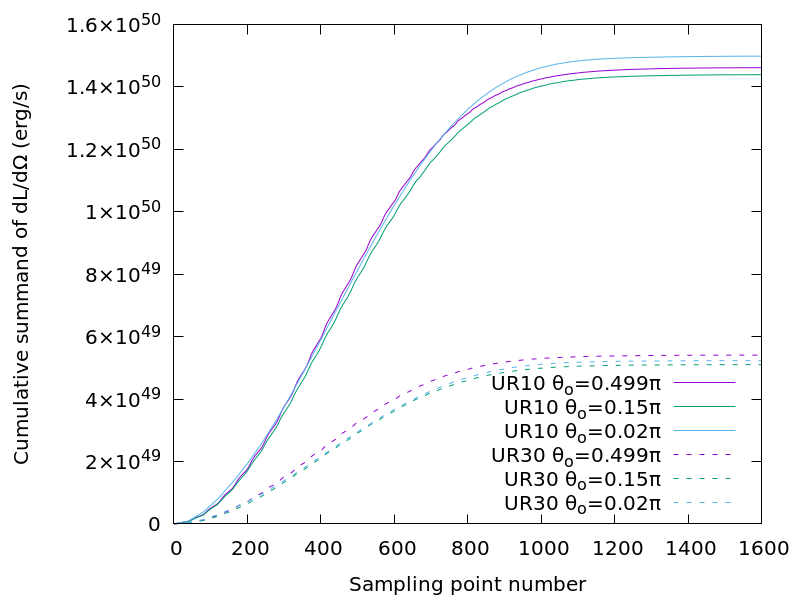}
    \caption{Cumulative distributions of the summands of \(dL/d\Omega\) for \(\nu_e\) in models UR10 (solid lines) and UR30 (dashed lines) for three different observer positions: the equatorial (purple), intermediate (green) and  polar (blue) observers.}
    \label{fig:cumulativesummands}
\end{figure}

It is difficult to judge from Fig. \ref{fig:summands} if the sum of all points is lowest for the intermediate-observer, thus producing the dip in Fig. \ref{fig:dLdO}. To confirm this, we present the cumulative distributions of the summands in Fig. \ref{fig:summands}. For this purpose, we assign sequential numbers to the sampling points, starting from the bottom point on the innermost circle and proceeding clockwise and from inside to outside; some of the numbers are shown in Fig. \ref{fig:summands}. We give in Fig. \ref{fig:cumulativesummands} the cumulative distributions of the summand (including the measure) of (the discretized version of) Eq. \ref{eq:obs-dependentLum} as a function of the number for the same models as in Fig. \ref{fig:summands}. 

At both times, the distributions for the intermediate-observer (green lines) are consistently the smallest of the three observers. This may be a bit counterintuitive, since there is no very reddish region for the pole-observer. This is a consequence of the fact that the central region is more bluish for the intermediate-observer, which is in turn owing to the angular distribution in momentum space, which is more inclined to the rotation axis at the corresponding point. This confirms the dip observed in Fig. \ref{fig:dLdO} at these times. The behavior of the cumulative distributions for the equator-  and pole-observers (purple and blues lines, respectively in Fig. \ref{fig:cumulativesummands}) is more complicated because of the appearance of the reddish regions at different places. The asymptotic value is greater for the pole-observer than for the equator-observer at \(t = 10\)s while the opposite is true at \(t = 30\)s. This is consistent with what we observe in Fig. \ref{fig:dLdO}. This trend, however, is not always true. At \(t=10\)s, for example, the trend is reversed at number \(\sim 750\) and once more at \(\sim 300\). The range in between corresponds indeed to the reddish region in the top left panel, where the equator-observer sees higher contributions to \(dL/d\Omega\). At \(t=30\)s, on the other hand, the reverse of the trend occurs only once at number \(\sim 200\), which corresponds to the inner edge of the reddish region for the equator-observer. This time the contributions from the outer points are not so high as at the earlier time for the pole-observer. 

Finally we mention the red dot that appears at a middle point in the triple crosses in the middle plots of Fig. \ref{fig:summands} for the intermediate-observer. It is isolated from the reddish region to the left. This is not an artifact. The dot corresponds to the north pole of the PNS. We find that the distribution function is largest there than in the surrounding area because of the same geometric effect. As a result, although the value of \(\theta_{\nu}\) for the observer direction is a bit larger on the red dot than at the lowest point in the triple crosses in the plots, the former contributes more to \(dL/d\Omega\) than the latter. Not to mention, the contribution from the uppermost point in the triple is smallest. It is noted that for the pole-observer (see the leftmost plots in Fig. \ref{fig:summands}) the north pole is located at the center and always bluish as explained above; for the equator-observer (the righmost plots), the north pole is barely seen at \(\theta_{\nu} = \pi/2\) and its contribution is negligible. This is why the red dot occurs only for the intermediate-observer. Even in that case, its contribution to \(dL/d\Omega\) is minor and does not prevent the dip from occurring for the intermediate-observer as observed in Fig. \ref{fig:dLdO}.

\subsection{Gravitational Waves}\label{Res-GW}

Since non-rotating spherically symmetric stars do not emit GWs, they are a good probe of asymmetry of the system. In this section, we evaluate the GWs emitted by the neutrinos that are radiated anisotropically themselves. For that purpose, we regard the two rotational sequences we have considered so far as the surrogate of the true temporal evolutions, which are currently unavailable. This is admittedly a very crude approximation but is certainly a substantial improvement to the previous estimates \cite{Fu_2022}, in which the asymmetry of the neutrino radiation from PNS was set by hand, and is the first step to more realistic evaluations of low-frequency GWs emitted by them. We employ the formulae explained in section \ref{NumMeth-GW}. 

As mentioned, since the PNS's and hence the neutrino radiations from them are axisymmetric, \(h_{\times}\) vanishes identically. In addition, the spherical-harmonic components with odd \(l\)'s also vanish due to the equatorial symmetry we impose in our models. In evaluating the GW spectra, we interpolate the snapshots and also extrapolate the resultant time evolution to \(50\)s.
In the exposition of the GW amplitudes, we assume that the observer is located at \(10\)kpc from the PNS.

\begin{figure*}
     \centering
     \begin{subfigure}[b]{0.45\textwidth}
         \centering
         \includegraphics[width=\linewidth]{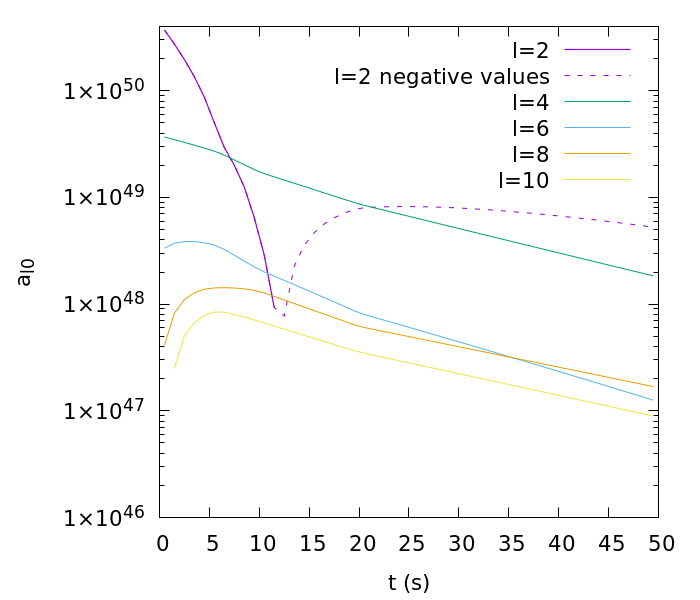}      
     \end{subfigure}
     \begin{subfigure}[b]{0.45\textwidth}
         \centering
         \includegraphics[width=\linewidth]{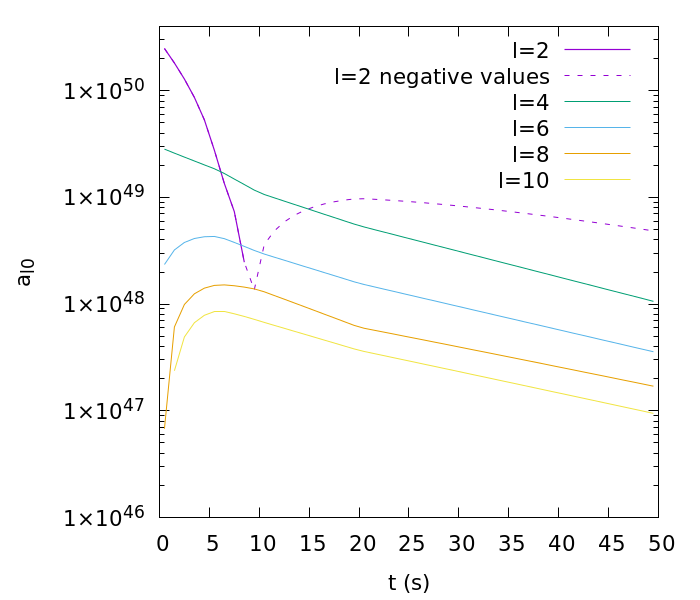}
     \end{subfigure}
        \caption{Time evolution of the spherical-harmonics expansion coefficients of the rigid-rotation model (left panel) and the differential-rotation model (right panel), including the contribution of the different \(l\)s.}
        \label{fig:GW-al0vst}
\end{figure*}

\begin{figure*}
     \centering
     \begin{subfigure}[b]{0.45\textwidth}
         \centering
         \includegraphics[width=\linewidth]{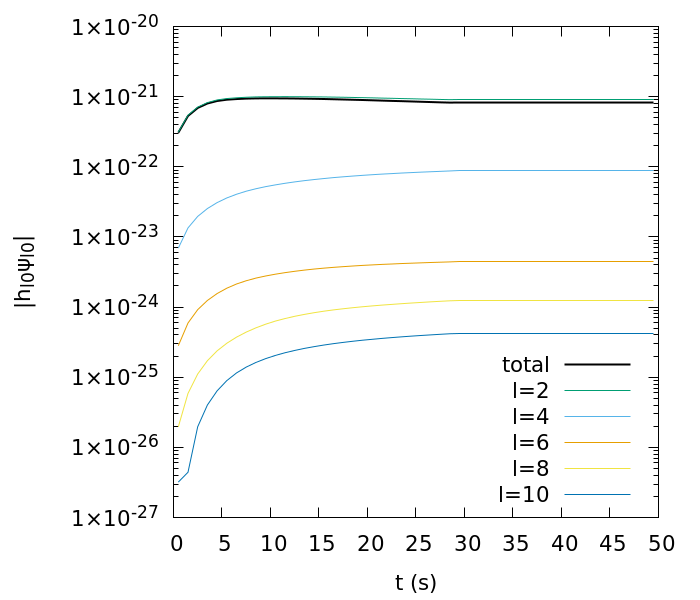}      
     \end{subfigure}
     \begin{subfigure}[b]{0.45\textwidth}
         \centering
         \includegraphics[width=\linewidth]{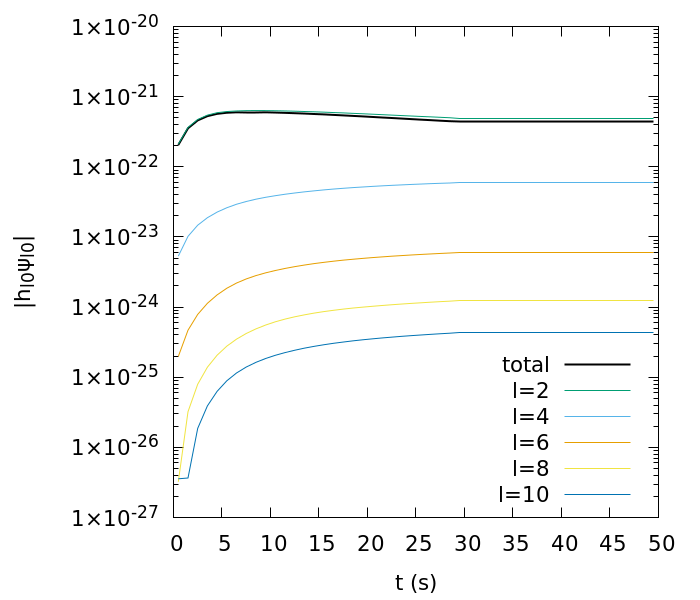}
     \end{subfigure}
        \caption{Waveform of the rigid-rotation model (left panel) and the differential-rotation model (right panel), including the contribution of the different \(l\)s.}
        \label{fig:GW-hPSIvst}
\end{figure*}

In order to understand better the behavior of the GW emissions, we begin with the study of the time evolution of the coefficients, \(a_{l0}\), in the spherical-harmonics expansion for the rigid- and differential-rotation models in Fig. \ref{fig:GW-al0vst}. The \(l=2\) contribution is dominant except around \(10\)s when it changes sign. This corresponds to the fact that the equator becomes dominant over the pole in \(dL/d\Omega\) around this time for both models (see Fig. \ref{fig:dLdO}). All the coefficients decrease in time but the dominant \(l=2\) component decays much more slowly than higher \(l\) components. Although the two rotational models are not much different from each other, a closer inspection finds that the rigid-rotation model yields higher values than the differential-rotation model, which results in stronger GW emissions for the former (see Fig. \ref{fig:GW-hPSIvst}). This is mainly because the former model rotates more rapidly at large radii and  its global asymmetry is larger.

Figure \ref{fig:GW-hPSIvst} shows the GW amplitudes as functions of time for the rigid- and differential-rotation models. In addition to their sum, the contributions from individual spherical-harmonics components are presented up to \(l = 10\). As expected from Fig. \ref{fig:GW-al0vst}, the contribution from \(l=2\) is dominant for both models. As the rigid-rotation models are more deformed, their GW amplitudes are a bit higher. This is also true for the individual spherical-harmonics components. Note that the total amplitude is slightly smaller than the \(l=2\) contribution, since the latter changes sign at \(t\sim 10\)s. For \(t\gtrsim30\)s, the GW amplitudes for both models approach non-vanishing values asymptotically. This is the memory effect.

\begin{figure*}
     \centering
     \begin{subfigure}[b]{0.45\textwidth}
         \centering
         \includegraphics[width=\linewidth]{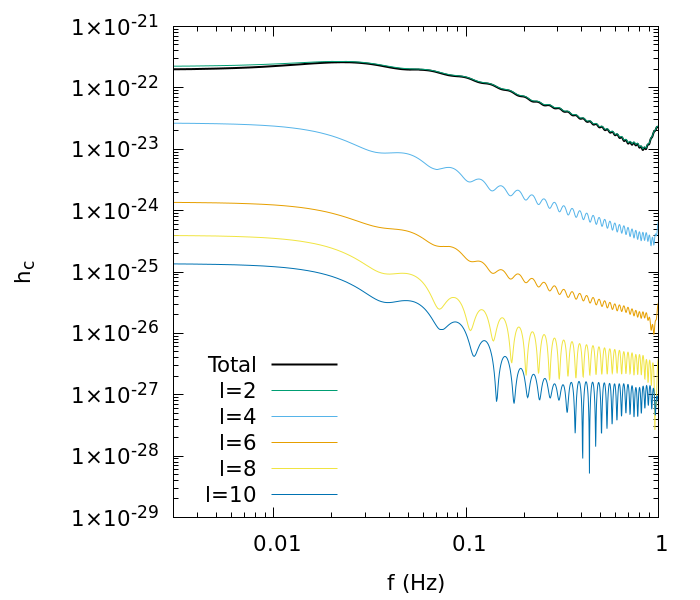}      
     \end{subfigure}
     \begin{subfigure}[b]{0.45\textwidth}
         \centering
         \includegraphics[width=\linewidth]{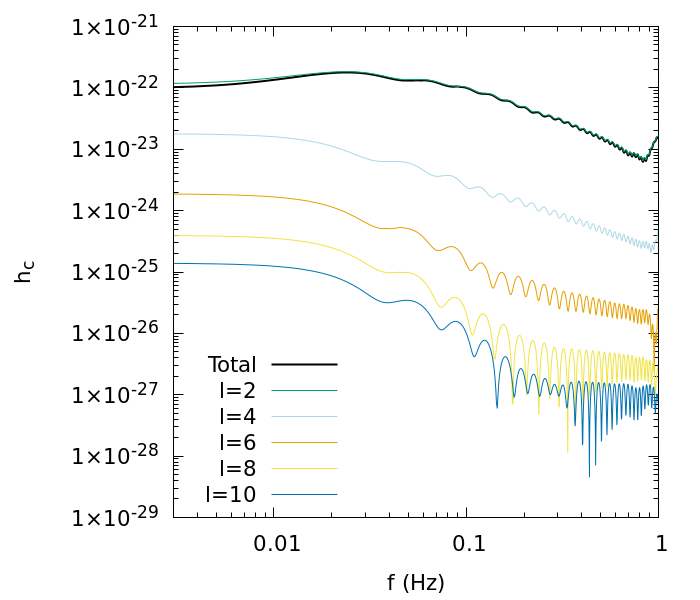}
     \end{subfigure}
        \caption{Characteristic strains for the rigid-rotation model (left panel) and the differential-rotating model (right panel), including the contribution of the different \(l\)s.}
        \label{fig:FFT}
\end{figure*}

The characteristic strains and their decompositions into the spherical-harmonics components are shown in Fig. \ref{fig:FFT}. As in the previous plots, the rigid-rotation model shows slightly larger characteristic strains than the differential-rotation model but the overall features are almost identical for the two rotational models with the contribution of \(l=2\) being dominant over other harmonics. The GW memory manifests itself here as non-vanishing of the characteristic strain at \(f \rightarrow 0\) (see Eq. (29) in \cite{Fu_2022}). It is found that the GW amplitudes in the  deci-Hz range obtained for our models are consistent with those derived in \cite{Fu_2022} with a more crude approximation. As claimed in \cite{Fu_2022}, we may read out the characteristic cooling time from the frequency, at which the GW starts to deviate from the asymptotic value at \(f \rightarrow 0\). In fact, the latter frequency is \(\sim0.03\)Hz for the dominant \(l=2\) component, which roughly corresponds to \(t_{\mathrm{cool}} \sim 5\)s. The cooling timescales derived from the temporal evolution of \(dL/d\Omega\) are given in Table \ref{tab:tcool}. They are obtained by fitting the five data points of \(dL/d\Omega\) for each observer direction with a continuous piecewise-linear function in the log-linear scale. One finds that the timescale obtained from the GW amplitude corresponds to the shortest timescale in the  earliest period. Since the neutrino luminosity is highest in the same interval, it is reasonable that the cooling timescale in that period is most remarkable. Although closer inspections give a hint of the longer timescales, particularly in higher harmonics, they will not be detectable observationally. Incidentally, the rise at  \(f\sim1\)Hz is an artifact of the temporal interpolation. Aside from this artifact, the detailed waveform (and hence the spectrum also) is affected by the interpolation scheme employed. Since the data are very sparse and the interpolant we choose, i.e., the piecewise-linear function in the log-linear scale, is one of the simplest, all the features at \(f \gtrsim 0.1\)Hz that could arise from the periods between the data points are simply ignored. They will be superimposed, if any, on the features we observe in Fig. \ref{fig:FFT}. We have to wait for multi-dimensional PNS-cooling calculations with rotation (and the evolution of the angular momentum distribution) fully taken into account to address this issue reliably. We are planning to do it in the near future.

\begin{table}
    \centering
    \begin{tabular}{|c|c|c|}
         \hline
         Time range (s) & \(t^{UR}_{\mathrm{cool}}\) range (s) & \(t^{DR}_{\mathrm{cool}}\) range (s) \\
         \hline
         \(2-6\) & \(6.2-7.1\) & \(6.5-7.2\)   \\
         \( 6-10\) & \(7.7-8.1\) & \(7.6-8.0\)  \\
         \( 10-20\) & \(14.6-15.3\) & \(14.6-15.2\)   \\
         \( 20-30\) & \(19.4-20.0\) & \(19.4-20.0\)   \\
         \hline
    \end{tabular}
    \caption{Cooling timescales obtained by the exponential fitting of the lightcurve for both the rigid- and differential-rotation models.}
    \label{tab:tcool}
\end{table}

\begin{figure}
    \centering
    \includegraphics[width=1\linewidth]{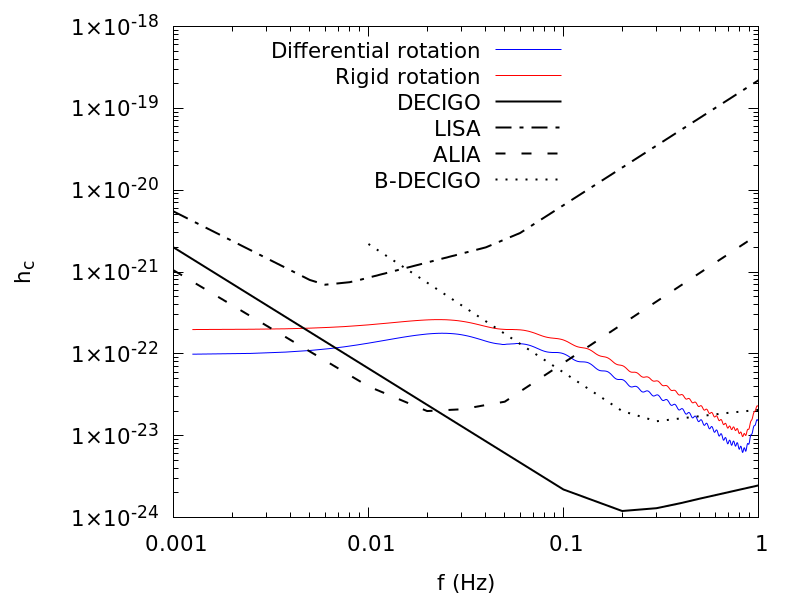}
    \caption{Sensitivity curves for LISA, DECIGO, ALIA and B-DECIGO compared with the characteristic strains of the rigid-rotation and differential-rotation models.}
    \label{fig:GW-FFT_det}
\end{figure}

Finally, we compare in Fig. \ref{fig:GW-FFT_det} the characteristic strains of both models with the planned sensitivities of some future detectors that are sensitive to the deci-Hz range of GWs \cite{Izumi_2021}: LISA \cite{Baker_2019}, DECIGO \cite{Yagi_2013}, ALIA \cite{Gong_2015} and B-DECIGO \cite{Isoyama_2018}. It is found that if the values predicted by our models are typical, the GW emissions from the cooling phase of a rapidly-rotating PNS at the Galactic center (\(D = 10\)kpc) can be detected by some of those upcoming detectors: whereas DECIGO can detect almost the whole signal from \(\sim 0.02\)Hz to \(\sim 1\)Hz, other detectors such as ALIA and B-DECIGO will be able to observe only a part of the range (\(\sim 0.01\) \(-\) \(\sim 0.1\)Hz and \(\sim 0.06\) \(-\) \(\sim 0.6\)Hz, respectively).

\subsection{Fast Flavor Conversion}\label{Res-FFI}
\begin{figure*}
     \centering
     \begin{subfigure}[b]{0.32\textwidth}
         \centering
         \includegraphics[width=\linewidth]{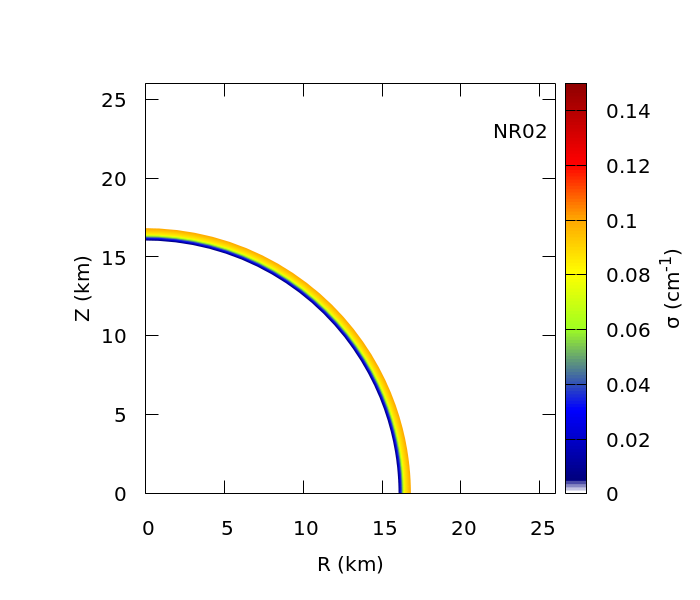}
     \end{subfigure}
     \begin{subfigure}[b]{0.32\textwidth}
         \centering
         \includegraphics[width=\linewidth]{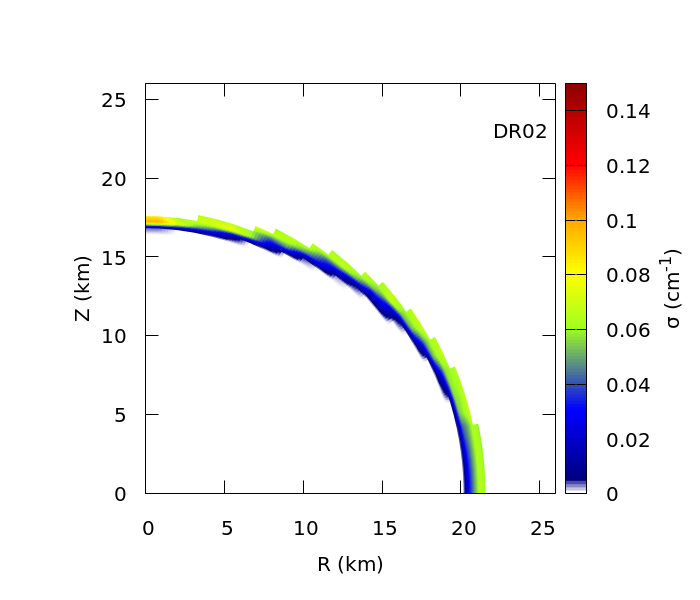}
     \end{subfigure}
     \begin{subfigure}[b]{0.32\textwidth}
         \centering
         \includegraphics[width=\linewidth]{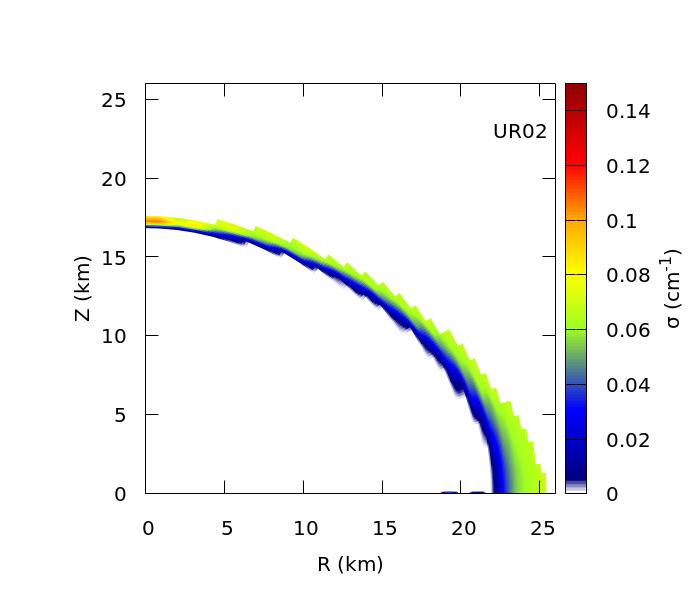}
    \end{subfigure}
     \centering
     \begin{subfigure}[b]{0.32\textwidth}
         \centering
         \includegraphics[width=\linewidth]{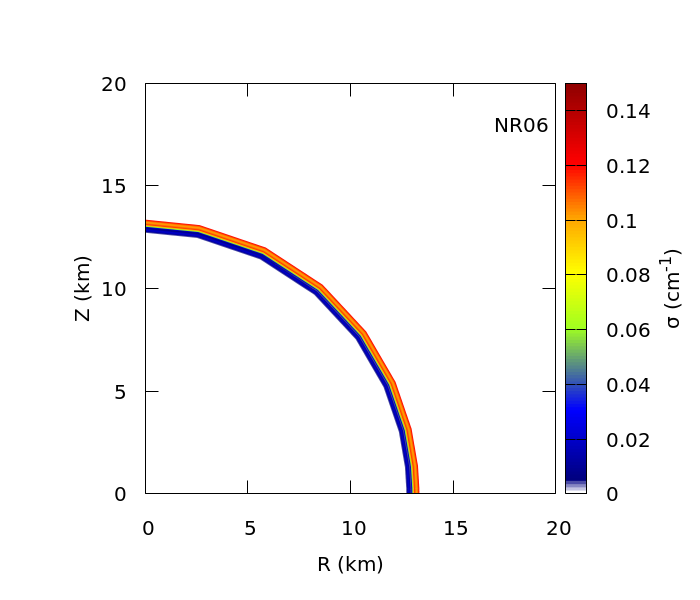}
     \end{subfigure}
     \begin{subfigure}[b]{0.32\textwidth}
         \centering
         \includegraphics[width=\linewidth]{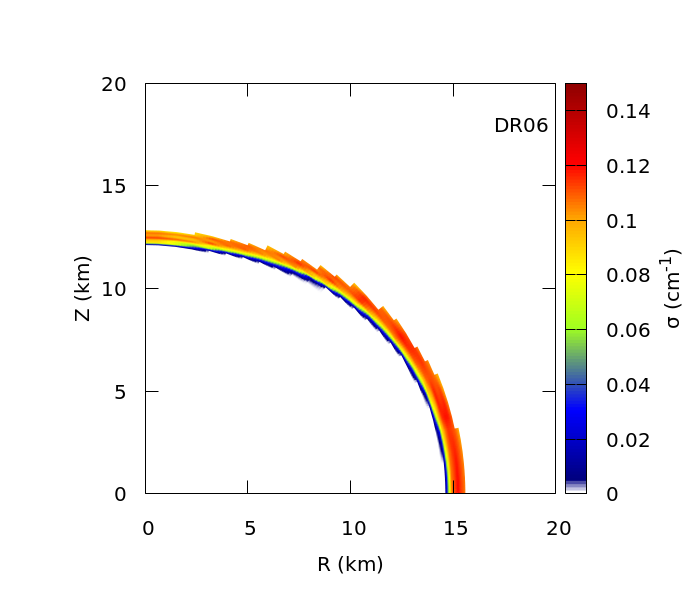}
     \end{subfigure}
     \begin{subfigure}[b]{0.32\textwidth}
         \centering
         \includegraphics[width=\linewidth]{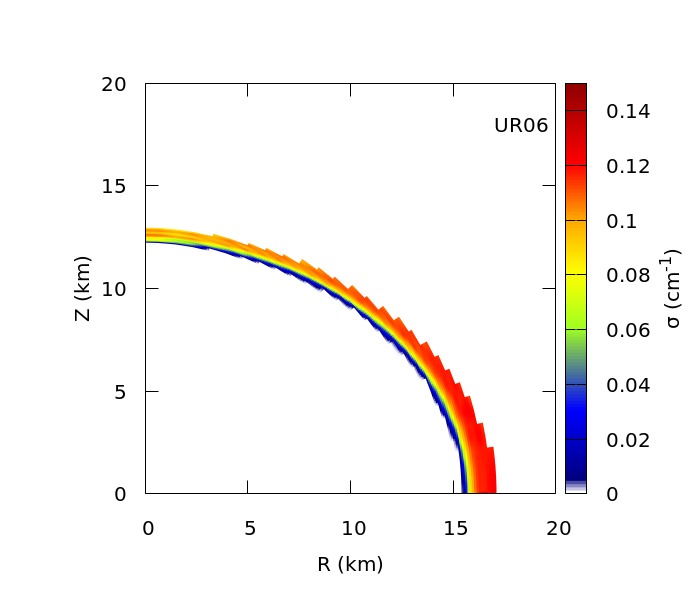}
     \end{subfigure}
     \centering
     \begin{subfigure}[b]{0.32\textwidth}
         \centering
         \includegraphics[width=\linewidth]{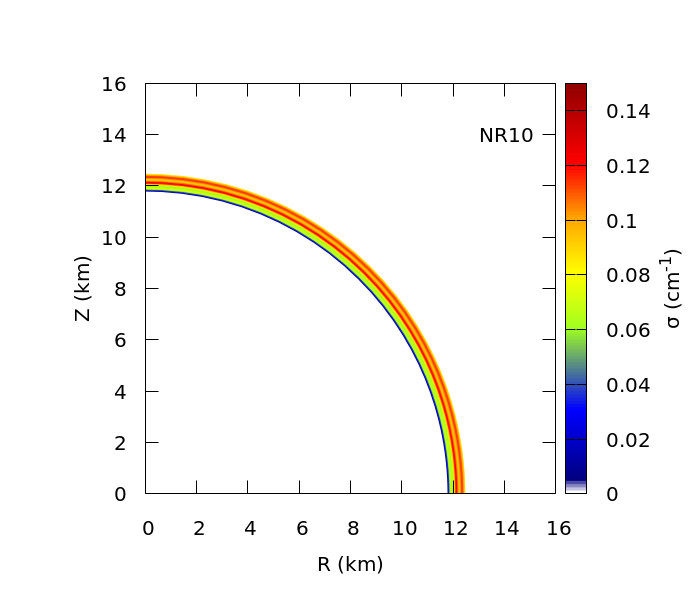}         
     \end{subfigure}
     \begin{subfigure}[b]{0.32\textwidth}
         \centering
         \includegraphics[width=\linewidth]{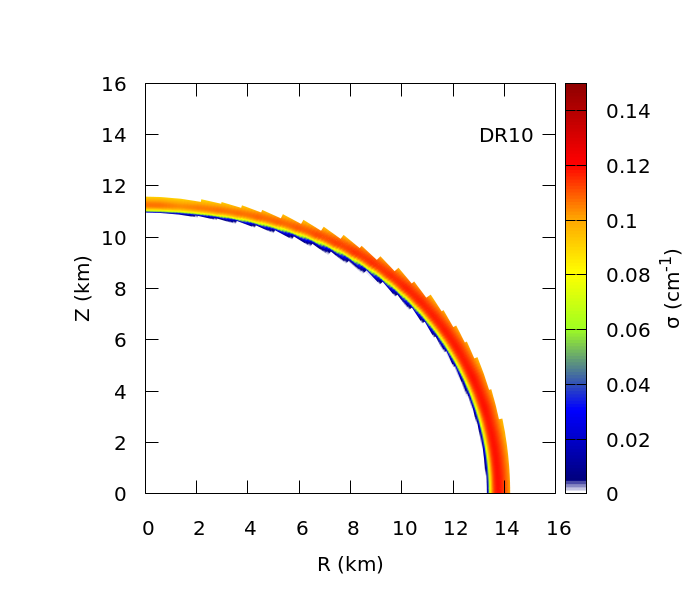}         
     \end{subfigure}
     \begin{subfigure}[b]{0.32\textwidth}
         \centering
         \includegraphics[width=\linewidth]{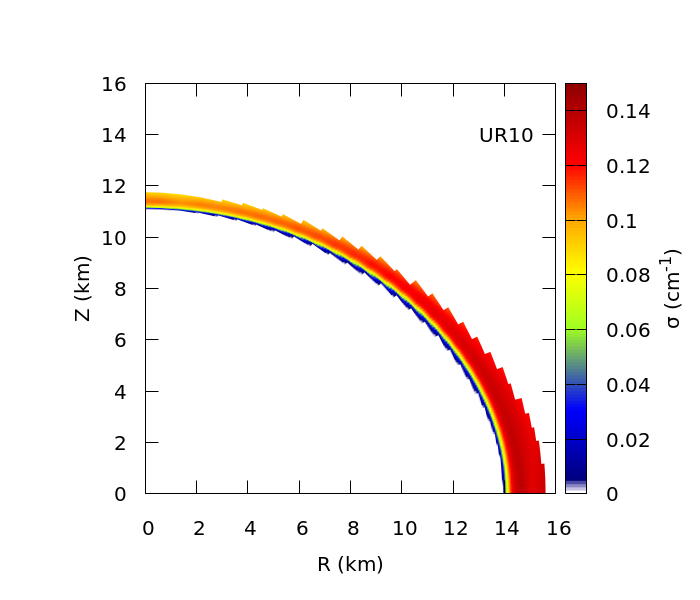}        
     \end{subfigure}
     \centering
     \begin{subfigure}[b]{0.32\textwidth}
         \centering
         \includegraphics[width=\linewidth]{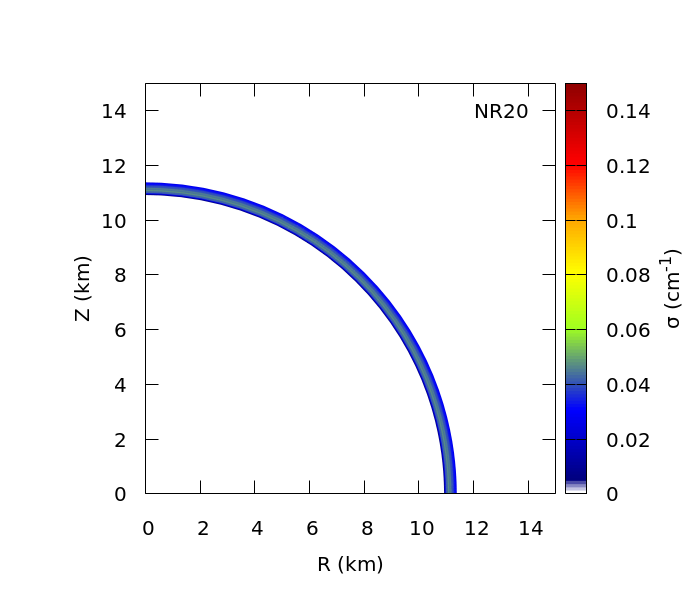}
     \end{subfigure}
     \begin{subfigure}[b]{0.32\textwidth}
         \centering
         \includegraphics[width=\linewidth]{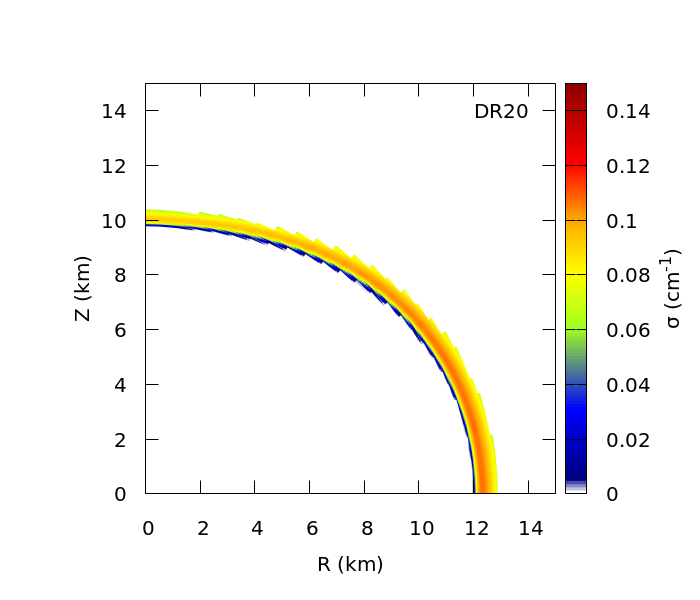}
     \end{subfigure}
     \begin{subfigure}[b]{0.32\textwidth}
         \centering
         \includegraphics[width=\linewidth]{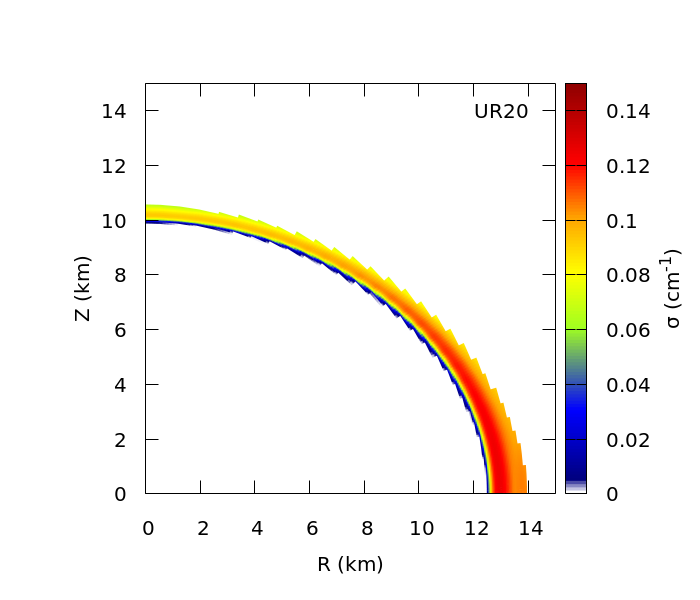}
     \end{subfigure}
     \centering
     \begin{subfigure}[b]{0.32\textwidth}
         \centering
         \includegraphics[width=\linewidth]{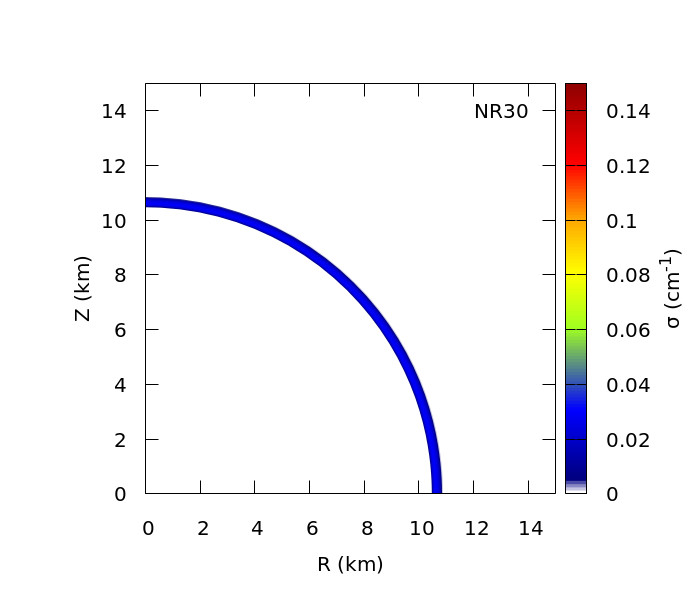}
     \end{subfigure}
     \begin{subfigure}[b]{0.32\textwidth}
         \centering
         \includegraphics[width=\linewidth]{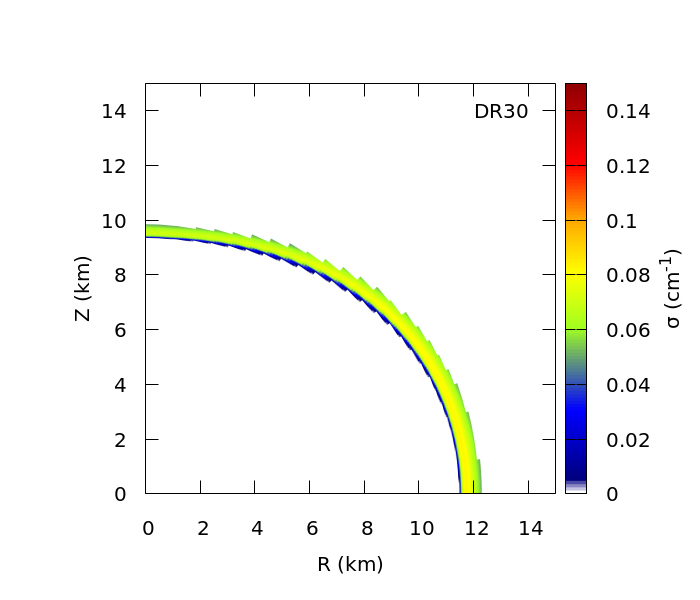}
     \end{subfigure}
     \begin{subfigure}[b]{0.32\textwidth}
         \centering
         \includegraphics[width=\linewidth]{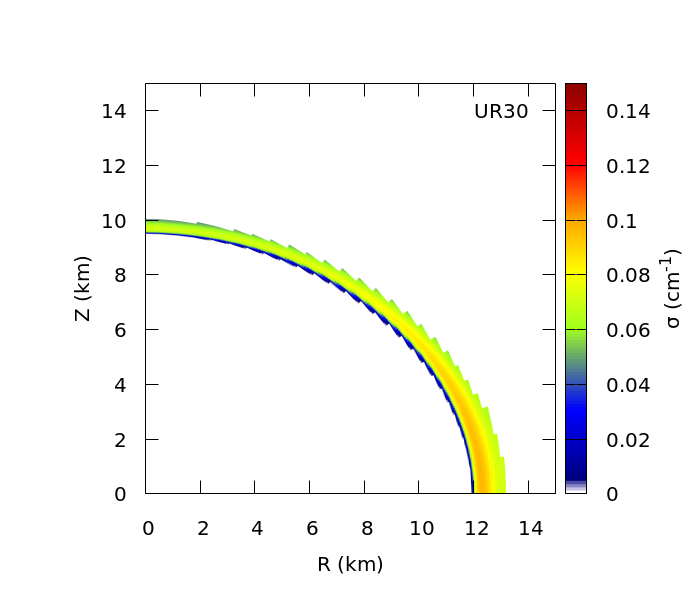}
     \end{subfigure}
        \caption{FFC growth rate in color in the quadrant of the meridian section for all the models.}
        \label{fig:FFI}
\end{figure*}

Finally, we touch the issue of collective neutrino-flavor conversions. In particular, we focus on the fast flavor conversion (FFC), since it is the most fast-growing mode if any. It is well known \cite{Zaizen_2024} that FFC happens between two flavor-sectors when one of them is more abundant in one angular domain while the other is dominant elsewhere in momentum space. It may be given as a zero-crossing of \(\Delta G\) in Eq. \ref{eq:DeltaG}, since we do not distinguish heavy-lepton neutrinos and their anti-particles. We take a post-process approach here, i.e., we look for the points that satisfy this condition in the results of simulations that ignore the flavor conversions, which is actually a common practice.
In Fig. \ref{fig:FFI}, we plot as color maps the (approximate) growth rate of FFC given by Eq. \ref{eq:sigmaFFI} for all models. In these plots, the region in white corresponds to no crossings and hence no FFC. As is apparent, there is a narrow region of FFC near the PNS surface for all the models at all times. This is in contrast to previous results \cite{Zaizen_2024}, in which no FFC region was observed in their spherically symmetric 1D models. The discrepancy may be due to the fact that while the previous studies employed mostly the multi-energy flux-limited diffusion approximation, we solve the Boltzmann equation faithfully; the neutrino-matter interactions are also handled differently. 

The regions, where we find crossings, are actually located outside but close to the neutrino-sphere and coincide with the regions, where both \(Y_e\) and \(Y_l\) rise with radius. This is why they are wider for the rotational models, particularly near the equator. In addition, the growth rate increases with time up to \(t=10\)s and then decrease thereafter, especially in the rigid-rotation model.

As mentioned at the beginning, these findings, particularly for the non-rotational models, seem at odds with the previous results. The regions are rather narrow and close to the stellar surface, at which we attach an extended tenuous atmosphere artificially for numerical reasons. Although we think that its influence is limited, we cannot exclude the possibility that our results are its artifact. Further investigations are certainly needed.

\section{Summary and Conclusions}\label{Sum}
We quantitatively evaluated the asymmetry in neutrino emissions from rapidly-rotating protoneutron stars (PNS's), employing the Boltzmann solver. We extracted snapshots at  \(t =2\), \(6\), \(10\), \(20\) and \(30\)s after a supernova explosion from a one-dimensional PNS cooling simulation without rotation and built two rotational equilibrium sequences by adding uniform or differential rotation by hand. They are both rapid rotators with \(T/|W| \sim 5 \% \). Having constant baryonic mass and total angular momentum, each sequence may be regarded as a surrogate for the actual time evolution. On top of the matter distributions so obtained, we solved numerically the Boltzmann equations for three neutrino species, \(\nu_e\), \(\bar{\nu}_e\) and \(\nu_x\), in (spatial) axisymmetry. From the results of these simulations we calculated the neutrino luminosity and the neutrino-hemisphere observer-wise. Interpolating and extrapolating the former in time, we also evaluated low-frequency (deci-Hz range) gravitational waves emitted by the neutrinos and studied their detectability. As the post-process of the simulation results that ignored neutrino oscillations entirely, we looked into the possibility of the fast flavor conversion. Our findings are summarized as follows.
\begin{itemize}
\item Our rotational models are both rapid rotators, having the rotational period of \(\sim 5\)ms on the equatorial surface initially, and are spun up further as they cool and shrink. As a result, they are oblate with the ratio of polar to equatorial radii of about \(0.7\) - \(0.8\), which is rather constant in time. Since the uniform-rotation model rotates faster at large radii, it is more flattened than the differential-rotation model.
\item Neutrinos are emitted anisotropically from the rotational PNS's. The luminosity that should be observed by a distant observer was calculated as a function of the observer position. It is highest for the polar observer and decreases monotonically toward the equatorial observer at the earliest time. As the time passes, however, there occurs a dip at an intermediate-observer direction. In addition, the luminosity becomes higher for the equatorial observer than for the polar observer at later times.
\item The neutrino-hemisphere, which was defined to be a set of points in PNS that have the optical depth of \(2/3\) from infinity, was calculated for different observer positions and the values of physical quantities such as neutrino mean energy on those points were investigated. The neutrino-hemisphere shrinks with time just as PNS does. It is not a perfect hemisphere with a constant radius even for the non-rotating case. It is skewed for the rotational cases, reflecting their oblate configurations. The oblateness of the neutrino-hemisphere is reduced in time in general. The mean neutrino energy on the neutrino-hemisphere shows a limb darkening. Interesting as they may be for their own sake, these effects do not explain the dip in the observer-dependent neutrino luminosity mentioned above.
\item The detailed investigation of the observer-dependent luminosity revealed that the dip was a product of the combination of geometric and rotational effects. The oblate PNS emits neutrinos in the direction of the rotation axis  more preferentially. As a result, the neutrino angular distribution in momentum space is inclined toward the axis. The misalignment is greatest at the intermediate-observer direction. On the other hand, rapid rotation of PNS makes the angular distribution inclined to the azimuthal direction, corresponding to the fact that neutrinos carry angular momentum. Both effects combined generates the dip.
\item The anisotropically-radiated neutrinos emit in turn gravitational waves in the deci-Hz range, which is set essentially by the cooling timescale. The spherical-harmonic analysis demonstrated that the \(l=2\) component is always dominant but changes sign at \(t \sim 10\)s. The contributions of higher harmonics get smaller monotonically with \(l\). The uniform- and differential-rotation models have similar behavior but the amplitudes are a bit higher for the former, since it rotates faster at large radii and is more deformed. The GW amplitude approaches a constant value asymptotically. The memory effect also manifests itself in the characteristic strain as a non-vanishing value at \(f \rightarrow 0\). It was found that the cooling timescale in the earliest phase can be read off from the spectrum. It was also demonstrated that the GWs from a Galactic event can be detected by some planned GW detectors such as DECIGO, B-DECIGO, LISA and ALIA in the deci-Hz range.
\item In contrast to the previous studies, there is a narrow region near the PNS surface in all our models including the non-rotational model that satisfies the linear instability condition and will undergo a fast flavor conversion. This may be due to the differences in the numerical methods employed but could be an artifact. Further investigations are needed.
\end{itemize}

There is admittedly an ample room for improvement. We did not compute the time evolutions of rotational PNS's but replaced them with a sequence of the rotational equilibrium configurations with prescribed rotation laws. As mentioned earlier, the angular momentum distribution in PNS is an outcome of the evolution itself and should be determined consistently. We simulated only neutrino transfer in this study, fixing matter motions (and the spacetime geometry). This is certainly a big flaw, since we are rather certain that there will be convective activities in PNS as it cools \cite{Roberts_2012,Nagakura_2020,Nagakura_2021,Akaho_2023}. If true, it may change the neutrino emissions even qualitatively. Some neutrino reactions of relevance particularly at later times \cite{Fischer_2020,Sugiura_2022,Fischer_2024} are neglected. These caveats notwithstanding, the current work is a step forward from the previous studies \cite{Mukhopadhyay_2021,Fu_2022}. We will address these issues one by one and one after another in the forthcoming papers.




\section*{Acknowledgements}
The authors want to thank Ken'ichi Sugiura for providing the background PNS profile. The author also owes Hirotada Okawa for advice and useful discussions. The author also gratefully acknowledges the help provided by Lei Fu regarding the GWs analysis.
This work used high performance computing resources provided by 
Cray XC50 at the National Astronomical Observatory of Japan (NAOJ), 
Fugaku supercomputer at RIKEN, the
FX1000 provided by Nagoya University, 
the Wisteria provided by JCAHPC
through the HPCI System Research Project (Project ID:
240041, 240079, 240219, 240264, 250006, 250166, 250191,
250226, JPMXP1020200109, JPMXP1020230406),
the Computing Research Center at the
High Energy Accelerator Research Organization (KEK),
Japan Lattice Data Grid (JLDG) on Science Information Network (SINET) of National Institute of Informatics (NII), Yukawa Institute of Theoretical Physics.
R. A. is supported by JSPS KAKENHI Grant-in-Aid for Scientific Research (24K00632). 
H. N. is supported by Grant-in-Aid for Scientific Research (23K03468). 
S. Y. is supported by Grant-in-Aid for Scientific Research (25K01006).

\nocite{*}

\bibliography{paper1}

\end{document}